\let\ifarxiv=\iftrue     % ARXIV VERSION
\documentclass[12pt,a4paper]{article}

\usepackage{amsmath,amssymb}
\usepackage[bookmarks=true,hyperfigures=true]{hyperref}
\usepackage{graphicx}
\usepackage[nosort]{cite}
\usepackage[a4paper,text={173mm,216mm},centering]{geometry}
\allowdisplaybreaks[3]

\numberwithin{equation}{section}

\usepackage{subfig}
\makeatletter
\let\old@startsection=\@startsection
\renewcommand{\@startsection}[6]{\old@startsection{#1}{#2}{#3}{#4}{#5}{#6\mathversion{bold}}}
\makeatother

\makeatletter
\newlength{\apb@width}
\newcommand{\autoparbox}[2][c]{\settowidth{\apb@width}{#2}\parbox[#1]{\apb@width}{#2}}

\makeatother

\let\oldPhi=\Phi
\let\oldPsi=\Psi
\let\oldGamma=\Gamma
\let\oldDelta=\Delta
\let\oldSigma=\Sigma
\let\oldLambda=\Lambda
\let\oldTheta=\Theta
\let\oldPi=\Pi
\let\oldXi=\Xi
\let\oldUpsilon=\Upsilon
\let\oldOmega=\Omega
\renewcommand{\Phi}{\mathnormal{\oldPhi}}
\renewcommand{\Psi}{\mathnormal{\oldPsi}}
\renewcommand{\Gamma}{\mathnormal{\oldGamma}}
\renewcommand{\Sigma}{\mathnormal{\oldSigma}}
\renewcommand{\Delta}{\mathnormal{\oldDelta}}
\renewcommand{\Theta}{\mathnormal{\oldTheta}}
\renewcommand{\Lambda}{\mathnormal{\oldLambda}}
\renewcommand{\Pi}{\mathnormal{\oldPi}}
\renewcommand{\Xi}{\mathnormal{\oldXi}}
\renewcommand{\Upsilon}{\mathnormal{\oldUpsilon}}
\renewcommand{\Omega}{\mathnormal{\oldOmega}}

\ifx\genfrac\sdflkaj\else\fi
\newcommand{\sfrac}[2]{{\textstyle\frac{#1}{#2}}}

\newcommand{\Tr}{\mathop{\mathrm{Tr}}}

\newcommand{\nn}{\nonumber}

\newcommand{\hi}{{\hat i}}   
\newcommand{\hj}{{\hat j}}   
\newcommand{\slsh}[1]{{#1}\!\!\!\! / \, }
\newcommand{\pslash}{p\!\! /}
\newcommand{\kslash}{k\!\! /}

\def\[{\begin{equation}}
\def\]{\end{equation}}
\def\<{\begin{eqnarray}}
\def\>{\end{eqnarray}}

\makeatletter
\def\mr@ignsp#1 {\ifx\:#1\@empty\else #1\expandafter\mr@ignsp\fi}%
\newcommand{\multiref}[1]{\begingroup%\let\protect\string%
\xdef\mr@no@sparg{\expandafter\mr@ignsp#1 \: }%
\def\mr@comma{}%
\@for\mr@refs:=\mr@no@sparg\do{\mr@comma\def\mr@comma{,}\ref{\mr@refs}}%
\endgroup}
\makeatother

\ifx\href\asklfhas\newcommand{\href}[2]{#2}\fi

\newcommand{\hypref}[2]{\ifx\href\asklfhas #2\else\href{#1}{#2}\fi}

\renewcommand{\eqref}[1]{(\multiref{#1})}

\newcommand{\tr}{{\rm Tr\,}}
\newcommand{\eqn}[1]{(\ref{#1})}
\ifx\hypersetup\sadfkjashdfkxja\newcommand\hypersetup[1]{}\fi

\hypersetup{plainpages=false}
\hypersetup{pdfpagemode=UseNone}
\hypersetup{bookmarksnumbered=true}
\hypersetup{pdfstartview=FitH}
\hypersetup{colorlinks=false}
\hypersetup{citebordercolor={.5 1 .5}}
\hypersetup{urlbordercolor={.5 1 1}}
\hypersetup{linkbordercolor={1 .7 .7}}
\begin{document}
\thispagestyle{empty}
%\ifarxiv\vspace*{-20mm}\fi

%\ifarxiv\else
\begingroup\raggedleft\footnotesize\ttfamily
%AEI-2010-019\\
HU-EP-10/14\\
%\arxivlink{1002.1733}
\par\vspace{15mm}
\endgroup
%\fi

\begingroup\centering
{\Large\bfseries\mathversion{bold} Light-like polygonal Wilson loops in 3d Chern-Simons and
ABJM theory\par}%
\ifarxiv\vspace{8mm}\else\vspace{15mm}\fi

\hypersetup{pdfauthor={Johannes Henn, Jan Plefka, Konstantin Wiegandt}}%
\begingroup\scshape\large 
Johannes M.~Henn,
Jan Plef\/ka
and Konstantin Wiegandt
\endgroup
\vspace{5mm}

\begingroup\ifarxiv\small\fi
\textit{Institut f\"ur Physik, Humboldt-Universit\"at zu Berlin, \\
Newtonstra{\ss}e 15, D-12489 Berlin, Germany}\par
\ifarxiv\texttt{\{henn,plefka,wiegandt\}@physik.hu-berlin.de\phantom{\ldots}}\fi
\endgroup

\ifarxiv\else
\vspace{5mm}

\begingroup\ttfamily
\verb+{henn,plefka,wiegandt}@physik.hu-berlin.de+\par
\endgroup
\fi

%\vspace{\fill}
\vspace{2cm}

\textbf{Abstract}\vspace{5mm}\par
\begin{minipage}{14.7cm}
We study light-like polygonal Wilson loops in three-dimensional Chern-Simons and ABJM theory to two-loop order. For both theories we demonstrate that the one-loop contribution to these correlators
cancels. For pure Chern-Simons, 
we find that specific UV divergences arise from diagrams involving two cusps, 
implying the loss of finiteness and topological invariance at two-loop order. Studying 
those UV divergences we derive anomalous conformal
Ward identities for $n$-cusped Wilson loops which restrict the finite part of the latter to  conformally invariant functions. 
We also compute the four-cusp Wilson loop in ABJM theory to two-loop order and find that the result is remarkably similar to that
of the corresponding Wilson loop in $\mathcal{N}=4$ SYM. Finally, we speculate about the
existence of a Wilson loop/scattering amplitude relation in ABJM theory.
\end{minipage}\par
\endgroup 
\newpage

\tableofcontents

%%%%%%%%%%%%%%%%%%%%%%%%%%%%%
%\pagenumbering{arabic}
%\setcounter{page}{1}
%\renewcommand{\thefootnote}{\arabic{footnote}}
%\setcounter{footnote}{0}

%\setcounter{tocdepth}{2}
%\hrule height 0.75pt
%\tableofcontents
%\vspace{0.8cm}
%\hrule height 0.75pt
%\vspace{1cm}

\setcounter{tocdepth}{2}

%%%%%%%%%%%%%%%%%%%%%%%%%%%%%%%%%%%%%%%%%%%%%%%%%%%%%%%%%%%%%%%%%%%%%%%%%%%%%%%%
%%%%%%%%%%%%%%%%%%%%%%%%%%%%%%%%%%%%%%%%%%%%%%%%%%%%%%%%%%%%%%%%%%%%%%%%%%%%%%%%
\section{Introduction and conclusions}

Wilson loops are the central non-local observables in any gauge theory and thus of
intrinsic interest. In 
3d Chern-Simons theory they are the principal observables and topologically invariant
with exactly known correlation functions \cite{Witten:1988hf} in the Euclidean (or Wick rotated) theory.
This exact result is analytic in the inverse Chern-Simons parameter $k$ 
and perturbative studies in a loop-expansion of the effective coupling constant $1/k$
can reproduce the exact topological and finite 
result to the first orders \cite{Guadagnini:1989am,Alvarez:1991sx}, modulo regularisation
subtleties leading to or not leading to an integer shift of $k$ (for a 
review see \cite{Labastida:1998ud}).  
Wilson loops in Minkowski-space with cusps and 
light-like segments, however, display particularly strong divergences in 4d gauge theories and
seem to not have been considered in the 3d Chern-Simons literature before.\\

In this paper we study such light-like Wilson loops with
cusps of polygonal shape in perturbation theory up to the next-to-leading order in $1/k$.
We do this for both the pure 3d Chern-Simons theory as well as its conformal
$\mathcal{N}=6$ supersymmetric extension known as Aharony, Bergman, Jafferis, Maldacena 
(ABJM) theory \cite{Aharony:2008ug}. The $\mathcal{N}=6$ ABJM theory is built upon an 
$SU(N)\times SU(N)$ gauge symmetry which allows for a planar $N\to\infty$ 
limit with $\lambda=N/k$ held 
fixed, where $k$ is the common absolute value of the Chern-Simons parameters 
of the two $SU(N)$ subgroups.
In this limit the ABJM theory is conjectured to be dual to type IIA string theory on $AdS_{4}\times\mathbb{CP}_{3}$, representing an exact
gauge-string duality pair very similar in nature to the well studied 4d $\mathcal{N}=4$ super 
Yang Mills/$AdS_{5}\times S^{5}$ string duality pair. Supersymmetric Wilson loops in ABJM theory have been defined in \cite{Drukker:2008zx,Chen:2008bp,Rey:2008bh} for the 1/6 BPS and recently in 
\cite{Drukker:2009hy} for the 1/2 BPS case. The
correlators for these loops of Euclidean, circular geometry are moreover known exactly
in terms of a supermatrix model \cite{Marino:2009jd} using localisation techniques \cite{Kapustin:2009kz}. 
Our motivation to consider polygonal light-like Wilson loops in
the 3d Chern-Simons ABJM gauge theory stems
from the Wilson loop/scattering amplitude duality in $\mathcal{N}=4$ super Yang-Mills.
This duality was discovered in the dual $AdS_{5}\times S^{5}$ string picture at 
strong gauge coupling in \cite{Alday:2007hr} 
and shown to exist also in the weak coupling regime \cite{Drummond:2007aua,Brandhuber:2007yx,Drummond:2007cf}
with profound consequences on the symmetries
of these correlators leading to a dual superconformal \cite{Drummond:2008vq} 
respectively Yangian symmetry \cite{Drummond:2009fd}
of scattering amplitudes, for reviews see \cite{Alday:2008yw,Henn:2009bd}. 
Moreover, there are many structural similarities of
the 3d $\mathcal{N}=6$ superconformal ABJM theory to $\mathcal{N}=4$ super Yang-Mills, most
notably the emergence of hidden integrability 
\cite{Minahan:2002ve,Beisert:2003yb, Beisert:2003tq, Bena:2003wd},
(for reviews see \cite{Tseytlin:2004cj,Belitsky:2004cz,Beisert:2004ry,Beisert:2004yq,Zarembo:2004hp,Plefka:2005bk,Minahan:2006sk,Arutyunov:2009ga})
in the planar limit
for the spectral problem of determining anomalous scaling dimensions of local operators
\cite{Minahan:2008hf,Gromov:2008qe}. \\

Given these insights the question arises whether there could also be such a 
scattering amplitude/Wilson loop duality in the ABJM theory. Scattering
amplitudes in the ABJM theory have been analysed by Agarwal, Beisert and McLoughlin 
\cite{Agarwal:2008pu} who in fact studied more general mass deformed superconformal 
Chern-Simons theories with extended supersymmetries at the one-loop order. 
There a vanishing result for the four-point one-loop amplitudes in the ABJM theory was found and
the authors speculated whether the two-loop scattering amplitudes in $\mathcal{N}=6$
Chern Simons (ABJM theory) could be simply related to the one-loop $\mathcal{N}=4$ Yang-Mills
amplitudes. The main result of our paper is that this picture is consistent at least up to
the two-loop order: We observe a cancellation of the one-loop graphs for null polygonal loops
and find that the four-cusp Wilson loop at the two-loop order is of the same functional
form as the four-point MHV amplitude of $\mathcal{N}=4$ up to constant numerical terms. \\

Specifically we calculate the expectation value of the $n$-cusped Wilson loop operator 
\begin{align}\label{eqn:Wilson-loop-operator}
\langle W_{n} \rangle :=  \frac{1}{N}
\, \langle 0 |\,
\tr \mathcal{P} \exp \left( i \oint_{\mathcal{C}} A_\mu dz^\mu \right)\, | 0\rangle
\end{align}
in the planar limit\footnote{I.e. we take the limit $N, k \rightarrow \infty$, $\frac{N}{k}$ finite.} for light-like polygonal contours $\mathcal{C}$ in pure Chern-Simons and ABJM theory, see Appendix \ref{app:Conventions} for conventions of the Lagrangian and the path ordering. \\

The contour of the $n$-sided polygon $\mathcal{C}$ is given by $n$ points $x_i$ $(i=1,...,n)$ 
where we parametrise each edge $\mathcal{C}_i$ via
\begin{align}\label{eqn:zi-definition}
z_i^{\mu}(s_i)= x_i^{\mu} + p_i^{\mu} s_i,\qquad \qquad p_i^{\mu} = x_{i+1}^{\mu} -x_i^{\mu}
\end{align}
where $s_i \in [\,0,1\,]$ and $\mathcal{C} = \mathcal{C}_1 \cup ... \cup\, \mathcal{C}_n$. 
The calculations are performed in 3-dimensional Minkowski space with metric
\begin{equation}
\eta_{\mu\nu} = \text{diag} (1,-1,-1)\,,
\end{equation}
and the segments of the contour are light-like, i.e. $p_i^2=0$. Note that due to the 
light-like contour there is no difference in ABJM theory between the standard loop operator 
\eqn{eqn:Wilson-loop-operator} above and the 1/6 BPS supersymmetric loop operator of 
\cite{Drukker:2008zx}, as the terms in the exponential coupling to the scalars
drop out. 

Wilson loops in Chern-Simons theory are usually defined with a framing procedure 
\cite{Witten:1988hf,Guadagnini:1989am,Alvarez:1991sx}
which may be thought of as a widening of the Wilson line to a ribbon.
This is necessary
in order to define an integer twisting number of the individual loop and acts
as a particular point-splitting regulator for collapsing gauge field propagators in perturbation theory
while preserving the topological structure of the theory. 
Here we refrain from framing our loops as we do not encounter the problem of 
collapsing gauge field propagators due to the piece-wise
linear structure of our loops. Moreover, the ABJM theory is not topological due to metric
dependent interactions in the matter sector, so that there is no need for framing from that
perspective either.
Instead we regulate our correlators by the method of dimensional reduction
which has been tested to the three-loop order in 
pure 3d Chern-Simons to yield a vanishing
$\beta$-function and to satisfy the Slavnov-Taylor identities \cite{Chen:1992ee}. 
Here the tensor algebra is performed in 3 dimensions to obtain scalar integrands and then 
the dimension of the integrations are analytically continued.\\

The outcome of our computations at one-loop order in pure Chern-Simons and ABJM theory is
that as claimed
\begin{equation}
\langle  W_4 \rangle_{\text{1-loop}} = 0\, \quad  (\text{analytically})\, ,\qquad
\langle  W_6 \rangle_{\text{1-loop}} = 0\, \quad (\text{numerically}) \,.
\end{equation}
Moreover, conformal Ward identities force $\langle W_{n}\rangle_{\text{1-loop}}$ 
to depend only on
conformally invariant cross ratios of the $(x_i-x_j)^{2}$ and for $n=4$ and $6$ we show that
these functions vanish. \\

At the two-loop order we computed the tetragonal Wilson-loop $W_{4}$ in pure Chern-Simons
and $\mathcal{N}=6$ superconformal Chern-Simons (ABJM theory). The result in dimensional
reduction regularisation with $d=3-2\epsilon$ for the correlator in pure planar 
Chern-Simons reads
\begin{align}
\langle W_4 \rangle^{\text{CS}} &= 1 -\left(\frac{N}{k}\right)^2  \left[\ln(2) \frac{(-x_{13}^2 \, \tilde \mu^2 )^{2\epsilon}}{2 \epsilon} + \ln(2) \frac{(-x_{24}^2 \, \tilde \mu^2 )^{2\epsilon}}{2 \epsilon} + \text{const.} \right]
+ {\cal O}[(\sfrac{N}{k})^{3}]\,.
\end{align}
We remark that this result displays a breakdown of finiteness and topological nature
of the light-like four-cusp Wilson loop in 3d Chern-Simons at the two-loop order 
%arising from a vertex graph
due to divergences\footnote{Regularisation 
is known to be a highly  subtle issue in 3d Chern-Simons theory. 
In particular one could compute this correlator using a framing procedure adapted to light-like
curves, which would be very interesting. We thank N.~Drukker for discussions on this point.}
associated to two cusps at a light-like distance, see section \ref{sect:vertex-diagram}.
For the same correlator in the ABJM theory we find
\begin{align}
\label{here}
\langle W_4 \rangle^{\text{ABJM}} &=1+ \left( \frac{N}{k}\right)^2 \left [- \left( \frac{(-{\mu^\prime}^2 \, x_{13}^2 )^{2\epsilon}}{(2\epsilon)^2} +\frac{(-{\mu^\prime}^2 \, x_{24}^2 )^{2\epsilon}}{(2\epsilon)^2} \right) + \frac{1}{2}  \ln^2\left(\frac{x_{13}^2}{x_{24}^2}\right) +\text{const.}
\right ] 
+ {\cal O}[(\sfrac{N}{k})^{3}]\, .
\end{align}
This is indeed of the same functional form as the one-loop result in $\mathcal{N}=4$ super
Yang-Mills theory, where one has \cite{Drummond:2007aua} 
%(for a review see \cite{Henn:2009bd})
\begin{align}
\langle W_4 \rangle^{\mathcal{N}=4\,\,\text{SYM}} &=1+ \frac{g^{2}N}{8\pi^{2}} \left [ -\left( \frac{(-{\mu}^2 \, x_{13}^2 )^{\epsilon}}{\epsilon^2} +\frac{(-{\mu}^2 \, x_{24}^2)^{\epsilon}}{\epsilon^2} \right) + \frac{1}{2}  \ln^2\left(\frac{x_{13}^2}{x_{24}^2}\right) +\text{const.}
\right ] 
+ {\cal O}[g^{4}N^{2}]\, . 
\end{align}
It would certainly be interesting to explore this relationship also beyond four cusps. Also
a two-loop computation of the four-particle scattering amplitudes in ABJM theory would be 
very desirable in order to compare to our result. \\

{}From the string perspective the scattering amplitude/Wilson loop duality in the
$AdS_{5}/CFT_{4}$ system arises from
a combination of bosonic and fermionic T-dualities under which the free 
$AdS_{5}\times S^{5}$ superstring is self-dual \cite{Berkovits:2008ic,Beisert:2008iq}.
Hence, for the existence of an analogue duality in ABJM theory one would 
require a similar self-duality of the $AdS_{4}\times \mathbb{CP}_{3}$ superstring under a
suitable combination of T-dualities. 
This problem was analysed by Adam, Dekal and Oz in \cite{Adam:2009kt} with a negative
outcome: The Green-Schwarz $\sigma$-model is \emph{not} 
self-dual under bosonic T-dualtities in the transverse $AdS_{4}$ directions combined
with fermionic ones
within the framework of the fermionic Buscher dualisation procedure employed in
\cite{Berkovits:2008ic}. However, the analysis of \cite{Adam:2009kt} started from a 
partially $\kappa$-symmetry gauge fixed formulation of the $AdS_{4}\times \mathbb{CP}_{3}$ superstring in terms of a supercoset $\sigma$-model 
\cite{Arutyunov:2008if,Stefanski:2008ik}.
This restriction was overcome in \cite{Grassi:2009yj} where, building upon a 
complete superspace formulation of $AdS_{4}\times \mathbb{CP}_{3}$
\cite{Gomis:2008jt}, again the non-existence of a
T-self-duality of the $AdS_{4}\times \mathbb{CP}_{3}$ superstring was found.
Interestingly however, a very recent paper \cite{Bargheer:2010hn} has uncovered a Yangian symmetry of tree-level amplitudes in ABJM theory pointing towards integrability of the latter.
Moreover, the authors argue about the possibility of a self-duality of the 
$AdS_{4}\times \mathbb{CP}_{3}$
superstring upon T-dualizing also along the $\mathbb{CP}_{3}$ directions, which could
provide a loop-hole for a scattering-amplitude/light-like Wilson loop duality of
the ABJM theory. In  order
to settle this question a two-loop scattering amplitude computation in ABJM theory would be very desirable in order to compare to our result \eqn{here}.\\

Our paper is organised as follows: In section 2 we perform the one-loop computation in pure Chern-Simons 
and ABJM theory for the tetragon and the hexagon Wilson loop. Section 3 is devoted
to the two-loop problem of the tetragon
in pure Chern-Simons theory where we compute all relevant 
graphs.  Then, in section 4, we perform an independent check of 
our results by deriving anomalous conformal
Ward identities for the Wilson loop. 
The latter can be generalised to an arbitrary number of points.
In section 5 we include the matter diagrams arising in
ABJM theory at the two-loop order and combine the results for the tetragon to obtain our
final result \eqn{here}. In the appendix we collect a detailed account of 
our conventions and give the technical details of the computation of the two-loop
graphs using Mellin-Barnes techniques.

\subsection*{Note added}
After publication of this paper the works \cite{Chen:2011vv}, \cite{Bianchi:2011dg} appeared which report on a two loop calculation of four-point scattering amplitudes in ABJM theory. Most interestingly, the results coincide precisely with the divergent and finite pieces of our Wilson loop computation in \eqref{here} up to a constant. In the published version of this article there was an erroneous sign in \eqref{eqn:result-matter-part} leading to a spurious sign difference between the Wilson loop and the scattering amplitude. We thank the authors of \cite{Bianchi:2011dg} for pointing this out to us.

%%%%%%%%%%%%%%%%%%%%%%%%%%%%%%%%%%%%%%%%%%%%%%%%%%%%%%%%%%%%%%%%%%%%%%%%%%%%%%%%
\section{One loop: Chern-Simons and ABJM theory}\label{sec:one-loop}
In this section we consider the one-loop expectation value of polygonal
Wilson loops with $n$ cusps. We would like to consider kinematical configurations for which all non-zero distances satisfy $-x_{ij}^2 >0 $, 
such that the result for the Wilson loops will be real (In particular, this allows us to
drop the $i \epsilon$ prescription of the propagators).
For $n$ odd, however, it is impossible to find vectors $p_{i}^{\mu}$
that lead to such configurations. For this reason, we will only discuss $n$ even.\\

At one-loop level, we only need terms quadratic in the expansion of the
Wilson loop operator, and the free part of the action. 
Therefore,  at one loop order, the expectation value of (\ref{eqn:Wilson-loop-operator}) 
in ABJM theory coincides with the one in pure Chern-Simons theory. 

\begin{figure}[t]
\centering
\subfloat[]{\begin{minipage}[c]{4cm}
		\includegraphics[width=0.85 \textwidth]{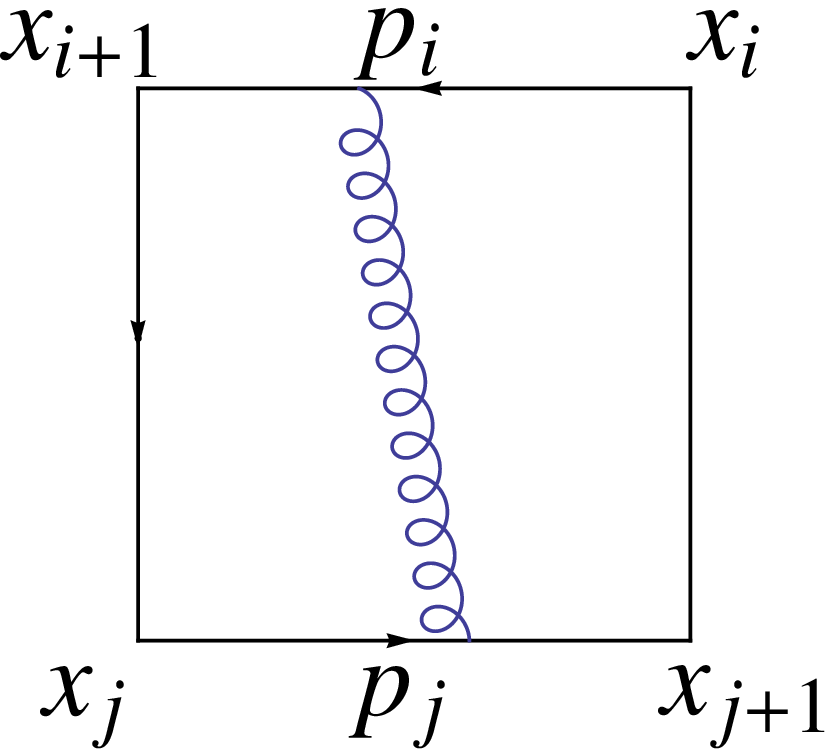}\label{fig:general-diagram}
		\end{minipage}
		
}
\subfloat[]{\begin{minipage}[c]{4cm}
		\includegraphics[width=0.6 \textwidth]{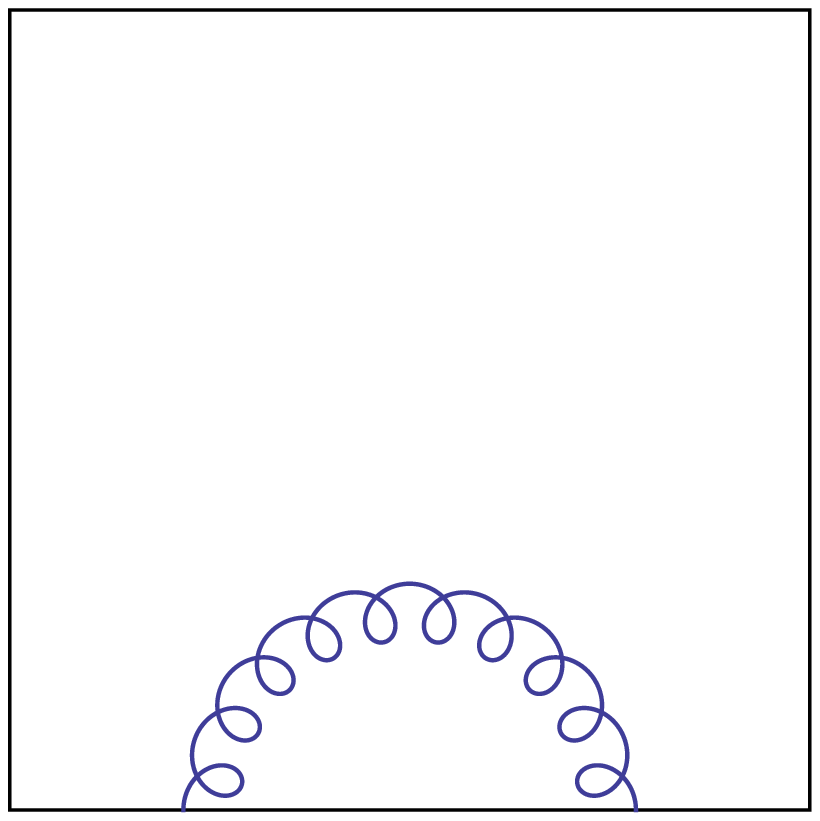}\label{fig: 1loopsameedge}
		\end{minipage}
		
}
\subfloat[]{\begin{minipage}[c]{4cm}
		\includegraphics[width=0.6 \textwidth]{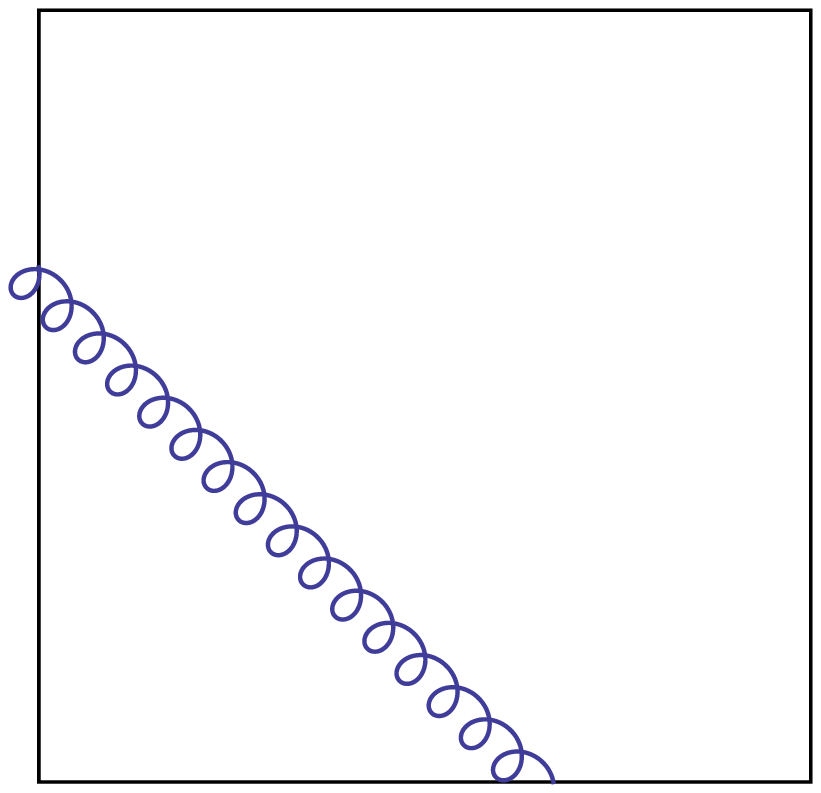}\label{fig: 1loopadjacentedge}
		\end{minipage}
		
}
\caption{
One-loop diagrams.
}
\label{fig:general-diagram-1-loop}
\end{figure}

The expectation value at one loop is a sum over all possible diagrams where the propagator stretches between edges $i$ and $j$,
\begin{align}\label{eqn:one-loop-general-expression1}
\langle W(\mathcal{C}) \rangle^{(1)}
&=  \frac{(i)^2}{N}\sum_{i \geq j}\int ds_i ds_j \dot{z}_{i}^{\mu} \dot{z}_{j}^{\nu} \langle \left(A_{\mu}\right)_{mn} (z_i)\left(A_{\nu}\right)_{nm}(z_j)\rangle
=:  -\frac{N}{k}  \frac{\Gamma\left(\frac{d}{2}\right) }{\pi^{\frac{d-2}{2}}} \, \sum_{i \geq j} I_{ij} \,,
\end{align}
where the domain of integration is given by $\int_0^1 ds_i \int_0^1 ds_j$ for $i \neq j$,  $\int_0^1 ds_i \int_0^{s_i} ds_j$ for $i = j$ 
and $\dot{z}(s_i)={d z(s_i)}/{d s_i}=p_i$ and where we have introduced a normalisation factor for later convenience.
Here and throughout the paper, we absorb the dimensional regularisation scale $(\mu^2 )^{\epsilon}$ into $k$ and only display it explicitly in our final results.
Using the Chern-Simons propagator  in the Landau gauge\footnote{We drop the $i \epsilon$ prescription in the propagator, since we consider kinematical configurations with $-x_{ij}^2>0$.}
\begin{align}\label{eqn:gluon-prop}
\langle \left(A_\mu\right)_{mn}(x) \left(A_\nu\right)_{kl}(y) \rangle &= \delta_{ml} \delta_{nk} \frac{1}{k}\left( \frac{\Gamma\left(\frac{d}{2}\right)}{\pi^{\frac{d-2}{2}}}\right) \epsilon_{\mu\nu\rho}\frac{(x-y)^\rho}{\left(-(x-y)^2\right)^{\frac{d}{2}}}\,,
\end{align}
and plugging in the expressions \eqref{eqn:zi-definition} for $z_i$,
we obtain 
\begin{align}\label{eqn:one-loop-general-expression}
I_{ij}= \int ds_i ds_j \frac{\epsilon(p_i,p_j, p_i s_i - p_j s_j + \sum_{k=j}^{i-1}p_k)}{\left(-x^2_{ij}\bar{s}_i\bar{s}_j - x^2_{i+1,j} s_i \bar{s}_j - x^2_{i,j+1}\bar{s}_i s_j-x^2_{i+1,j+1}s_i s_j\right)^{\frac{d}{2}}}\,,
\end{align}
where $x_{i,j}^2 = (x_i - x_j)^2$, $\bar{s}_i= 1- s_i$ and $\epsilon(a,b,c) = \epsilon_{ijk} a^{i} b^{j} c^{k}$.
We can immediately see that in this gauge $I_{i,i}$ and $I_{i,i+1}$ vanish due to the antisymmetry of the $\epsilon$ tensor.
This corresponds to diagrams where the propagator ends on the same edge or on adjacent edges, as shown in Figures \ref{fig:general-diagram-1-loop}(b) 
and  \ref{fig:general-diagram-1-loop}(c), respectively.
Therefore we only need to keep diagrams of the type shown in Figure \ref{fig:general-diagram-1-loop}(a). 
The latter are manifestly finite in three dimensions and therefore we set $d=3$ in the remainder of this section.

\subsection{Tetragon}
As explained above, in the Landau gauge, the 
only non-vanishing contributions to \eqref{eqn:one-loop-general-expression1} for the tetragon are $I_{31}$ and $I_{42}$.
Setting $d=3$, they are given by
\begin{align}
I_{31}=-\epsilon(p_1,p_2,p_3)\int ds_1 ds_3 \frac{1}{(-x^2_{13} \bar{s}_1\bar{s}_3 - x^2_{24} s_1 s_3)^{3/2}}\,, 
\end{align}
and
\begin{align}
I_{42}=-\epsilon(p_2,p_3,p_4)\int ds_2 ds_4 \frac{1}{(-x^2_{24} \bar{s}_2\bar{s}_4 - x^2_{13} s_2 s_4)^{3/2}}\,.
\end{align}
Taking into account that we have a closed contour, i.e. $\sum_i p_i =0$, we can write 
$\epsilon(p_2,p_3,p_4) = - \epsilon(p_2,p_3,p_1)= - \epsilon(p_1,p_2,p_3)$ and 
thus the contributions from the two diagrams cancel each other
\begin{align}
\langle W_4 \rangle^{(1)} & \propto \left( ~
\begin{minipage}[t]{50pt}
\vspace{-30pt}\includegraphics[height=50pt]{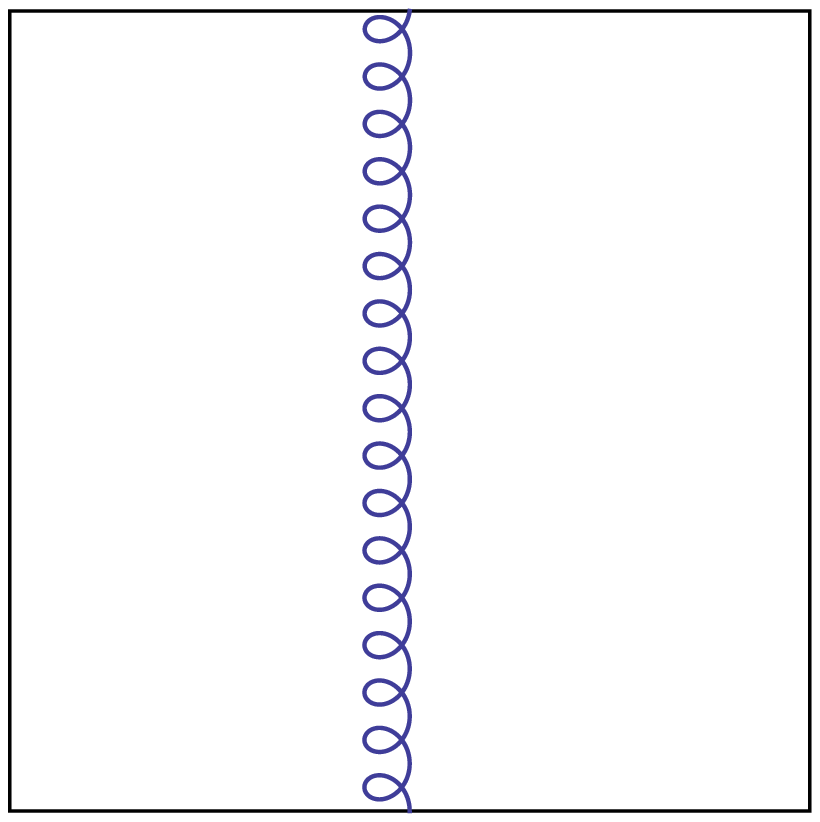} 
\end{minipage}
~~+~ 
\begin{minipage}[t]{50pt}
\vspace{-30pt}\includegraphics[angle=90,height=50pt]{i13} 
\end{minipage}~
\right) 
= I_{31} + I_{42} =0\,.
\end{align}
We will see in section \ref{sec:anomalous-ward-identities} that this result is compatible with the 
restrictions imposed by conformal symmetry.

\subsection{Hexagon and higher polygons}
\begin{figure}[t]
\centering
{\includegraphics[width=0.15 \textwidth]{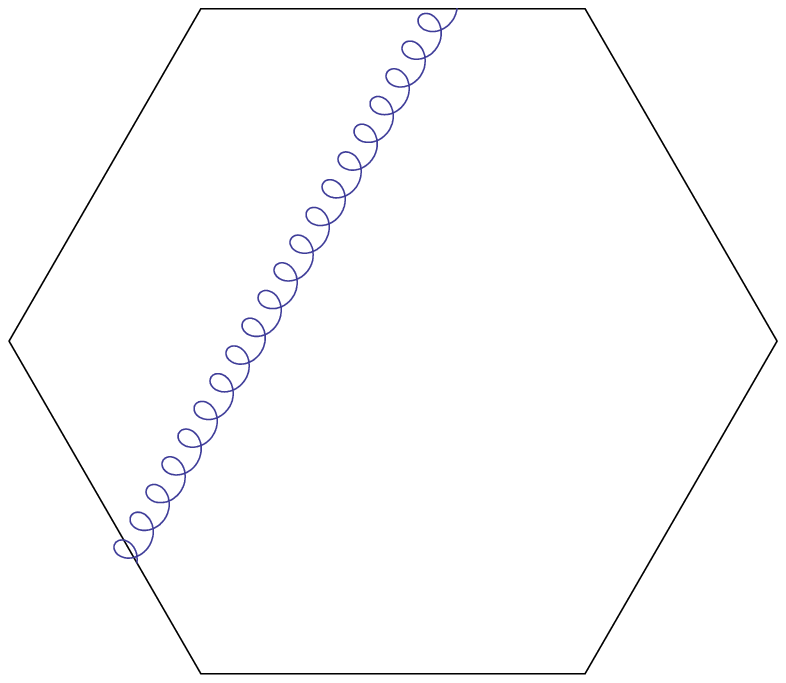}\label{fig:hexagon2}~~~~~~}
{\includegraphics[width=0.15 \textwidth]{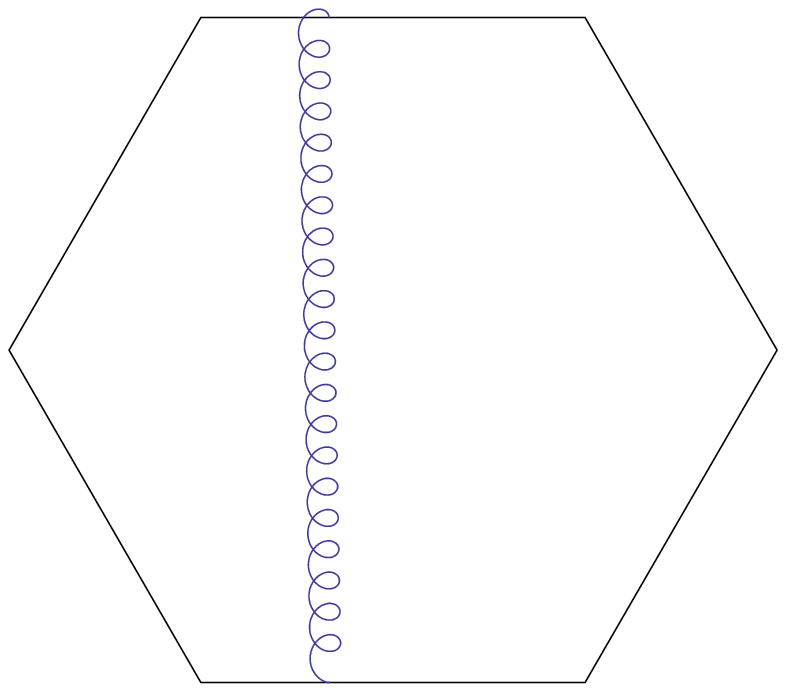}\label{fig:hexagon3}}
\caption{One-loop contributions $I_{i+2,i}$ and $I_{i+3,i}$ to the hexagonal Wilson loop.}
\label{fig:hexagon-one-loop}
\end{figure}

For the hexagon there are two different non-vanishing types of contributions, 
$I_{i+2,i}$ and $I_{i+3,i}$, as shown in Figure \ref{fig:hexagon-one-loop}. 
The former appears in six orientations, $i=1\, \ldots 6$ (with the convention that $i + 6 \equiv i$), while the
latter appears in three orientations, $i=1, 2, 3$. 

Specialising the general formula (\ref{eqn:one-loop-general-expression}) to these cases we have
\begin{equation}
\label{eqn:I_ii+2}
I_{i+2,i}=   \int_0^1 ds_{i+2} ds_{i} \frac{\epsilon(p_{i+2},p_i,p_{i+1})}{(-\bar{s}_i\bar{s}_{i+2} x_{i,i+2}^2 - s_{i}\bar{s}_{i+2} x_{i,i+3}^2 - s_i s_{i+2}x_{i+1,i+3}^2 )^{3/2}}\,
\end{equation}
and 
\begin{equation}
\label{eqn:I_ii+3}
I_{i+3,i}=  \int_0^1 ds_{i+3}ds_i \frac{\epsilon(p_{i+3},p_i,p_{i+1}+p_{i+2}) }{(-\bar{s}_i\bar{s}_{i+3} x_{i,i+3}^2 - s_{i}\bar{s}_{i+3} x_{i+1,i+3}^2 - \bar{s}_i s_{i+3}x_{i,i+4}^2 - s_i s_{i+3} x^2_{i+1,i+4})^{3/2} }\,.
\end{equation}

We checked numerically for various non-symmetric hexagon 
configurations that the sum over all diagrams vanishes,
\begin{align}\label{eqn:six-vanishes}
\langle W_6 \rangle^{\text(1)} \propto \sum_{i>j}^6 I_{ij} = 0\,.
\end{align}
Although we do not yet have an analytical proof for generic kinematical configurations,
we can show that \eqref{eqn:six-vanishes} is true for special configurations, as we will see presently.

Consider the configuration where opposite edges are anti-parallel, i.e. $p_i =- p_{i+3}$. 
{}From  \eqref{eqn:I_ii+3} we see that $I_{i,i+3}=0$ due to the antisymmetry of the $\epsilon$ tensor.
Furthermore, taking into account that for this configuration we have $x_{i,i+2}^2 = x_{i+3,i+5}^2$,
it is easy to see from equation \eqref{eqn:I_ii+2}  that the integrands of $I_{i,i+2}$ and $I_{i+3,i+5}$ are the same.
Finally, using $\sum_i p_i=0$ one can see that the Levi-Civita symbols produce a differing sign, such
that
\begin{align}
I_{i,i+2} + I_{i+3,i+5}&= 
\left(
\begin{minipage}{1.5cm}
{\includegraphics[width=1 \textwidth]{hexagon2}} 
\end{minipage} 
+ 
\begin{minipage}{1.5cm} 
{\includegraphics[width=1 \textwidth]{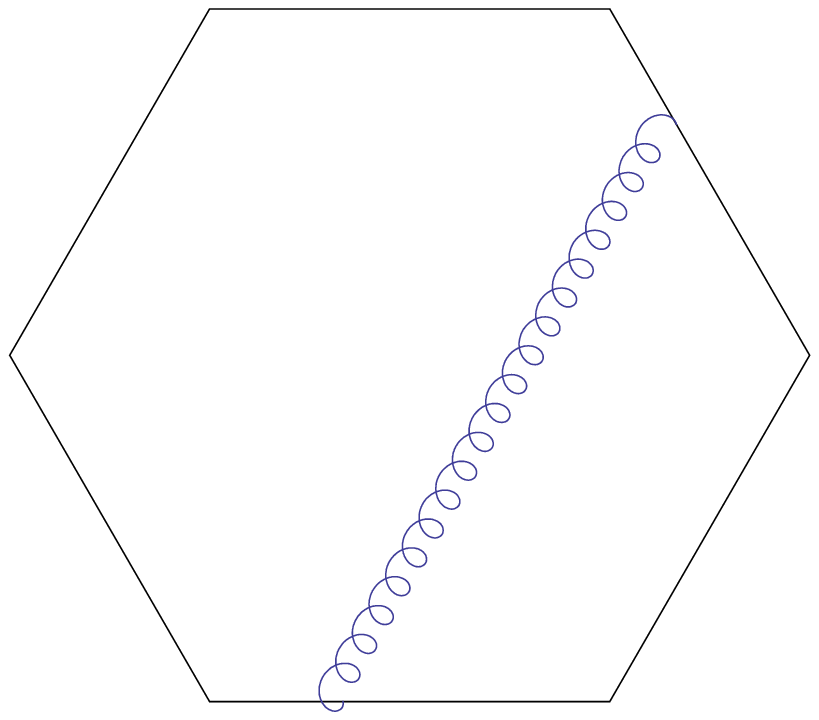}} 
\end{minipage}\right)_{p_i=-p_{i+3}} = 0 \,,
\end{align}
i.e. the contributions coming from those diagrams cancel pairwise,
and we arrive at equation  \eqref{eqn:six-vanishes}, in the specific anti-parallel kinematical configuration $p_i =- p_{i+3}$.

It is tempting to speculate that all $n$-cusped Wilson loops vanish at one-loop order in Chern-Simons theory.

%%%%%%%%%%%%%%%%%%%%%%%%%%%%%%%%%%%%%%%%%%%%%%%%%%%%%%%%%%%%%%%%%%%%%%%%%%%%%%%%

\section{Two loops: Chern-Simons theory}\label{sec:two-loop}

\begin{figure}[t]
\centering
\subfloat[]{\begin{minipage}{4cm}\centering
		\includegraphics[width=.6 \textwidth]{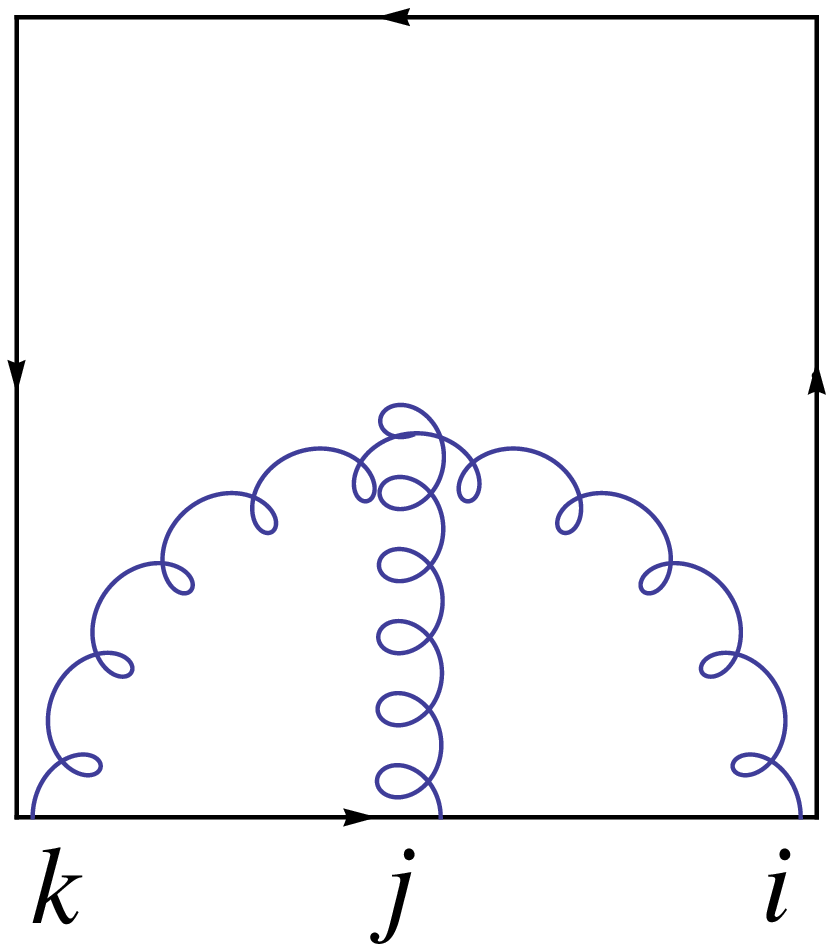}\label{fig:vertexI111}
		\end{minipage}
		
}
~~~~~
\subfloat[]{\begin{minipage}{4cm}\centering
		\includegraphics[width=.6 \textwidth]{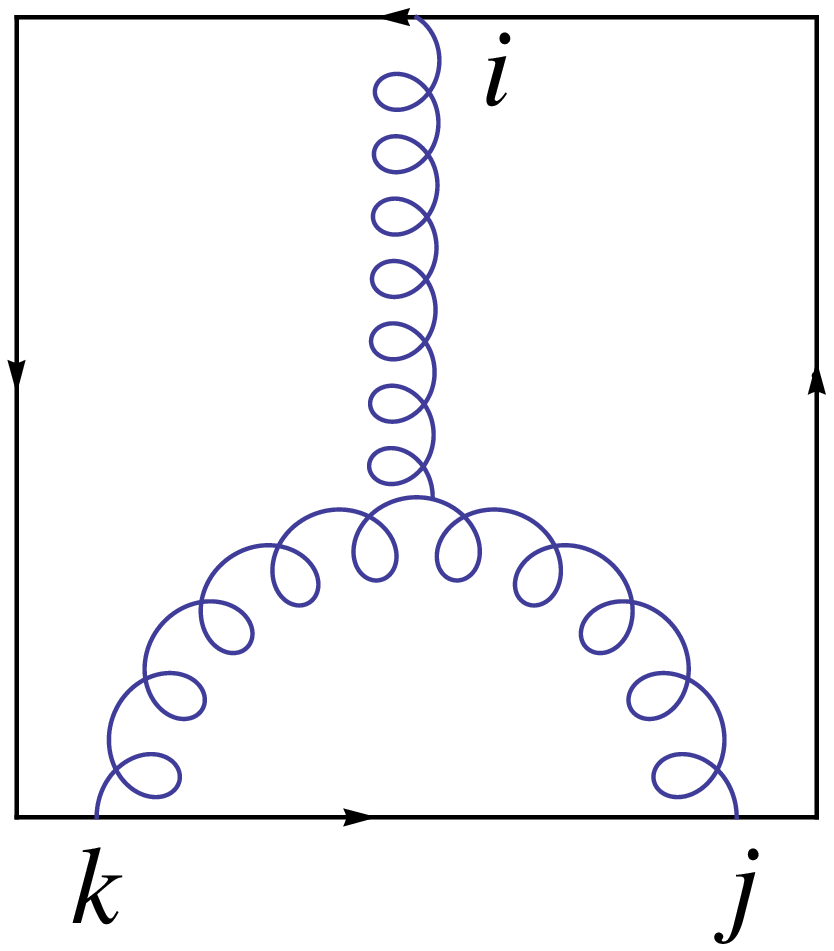}\label{fig:vertexI113}
		\end{minipage}
}
~~~~~
\subfloat[]{\begin{minipage}{4cm}\centering
		\includegraphics[width=.65 \textwidth]{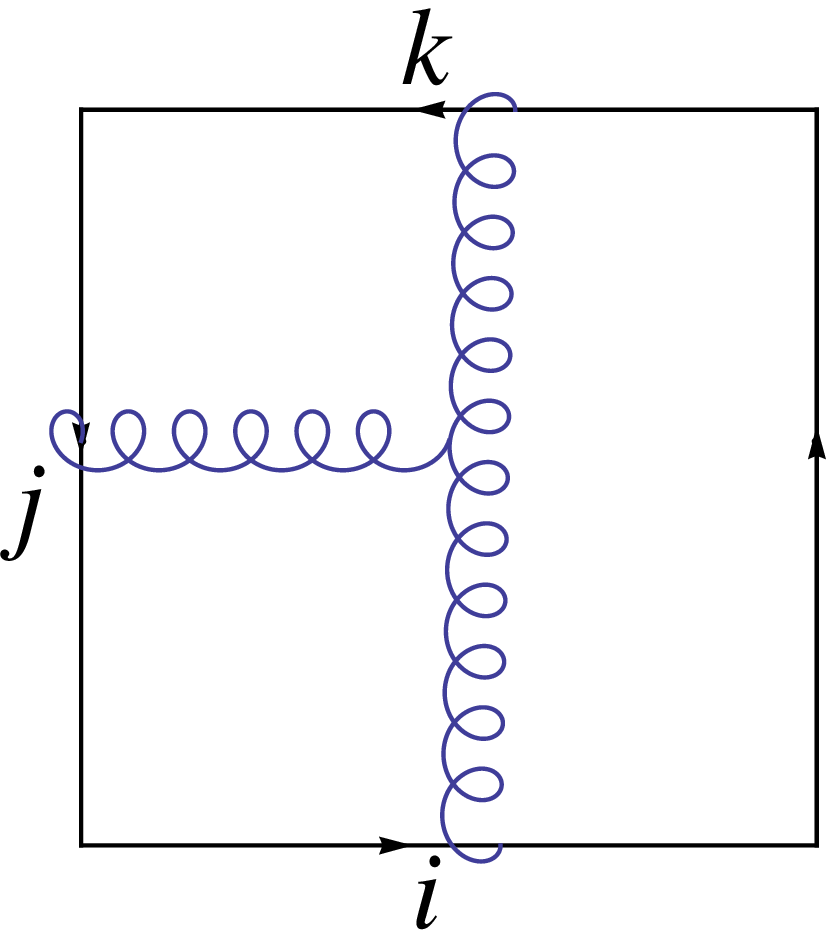}\label{fig:vertexI123}
		\end{minipage}
}\\

\subfloat[]{\begin{minipage}{4cm}\centering
		\includegraphics[width=.6 \textwidth]{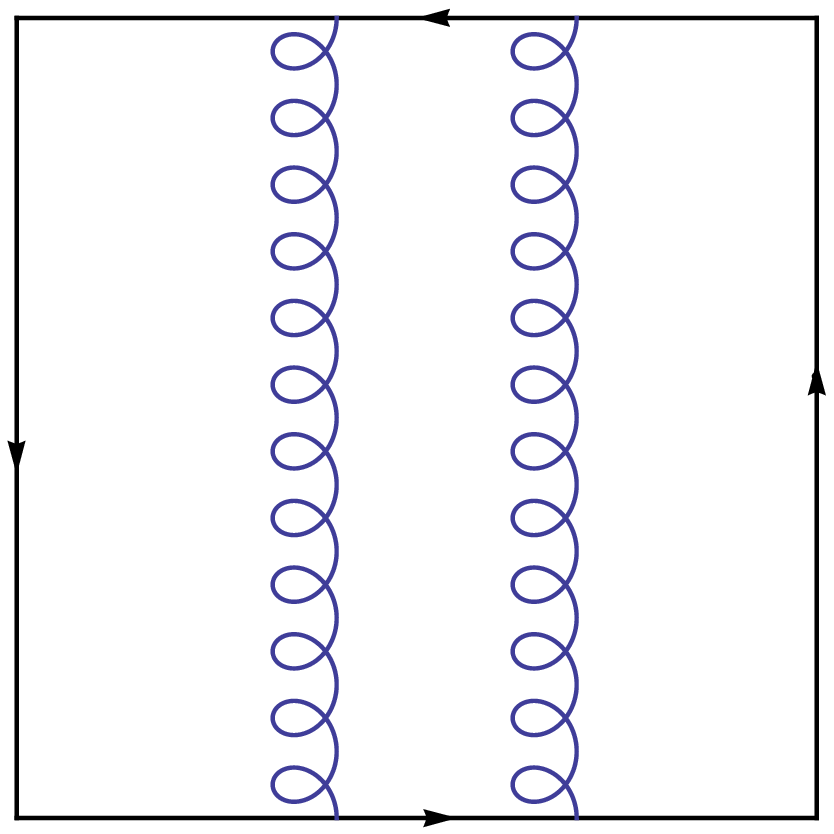}\label{fig: rect1331}
		\end{minipage}
}
\subfloat[]{\begin{minipage}{4cm}\centering
		\includegraphics[width=.6 \textwidth]{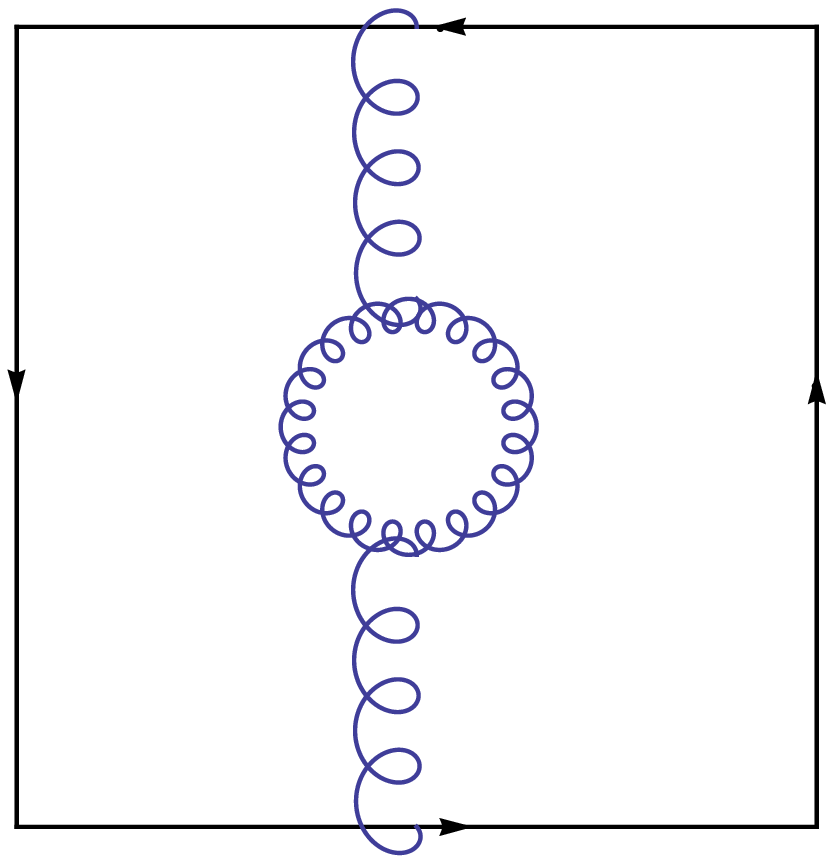}\label{fig:gluon-loop}
		\end{minipage}
}
\subfloat[]{\begin{minipage}{4cm}\centering
		\includegraphics[width=.6 \textwidth]{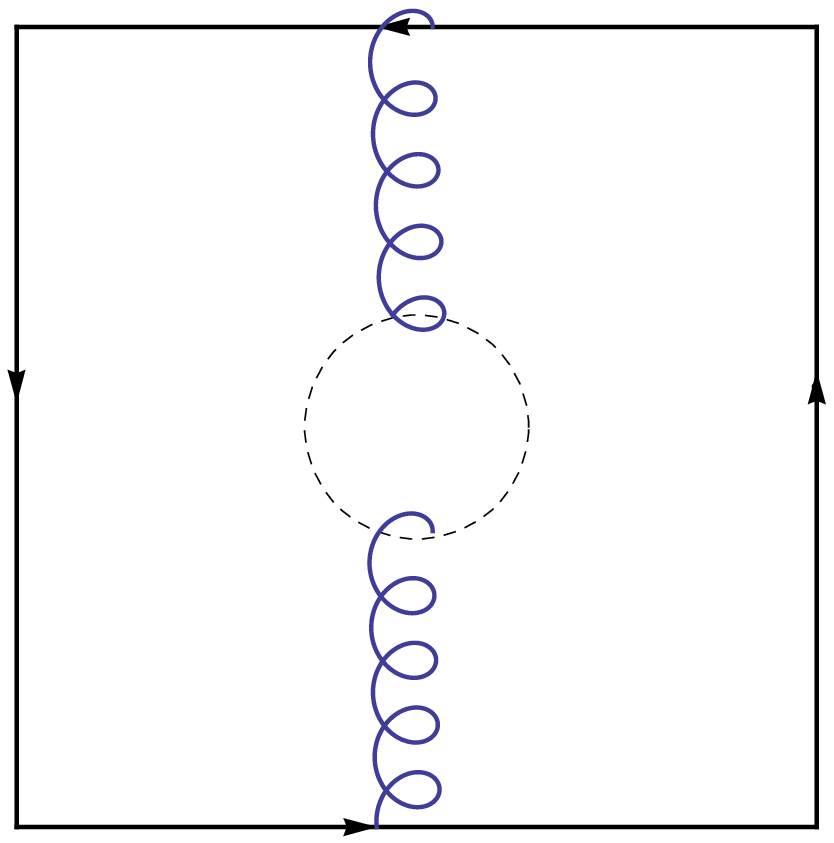}\label{fig:ghost-loop}
		\end{minipage}
}

\caption{Planar two-loop topologies appearing in the polygonal Wilson loop in CS theory.
Diagrams where one propagator is connected to a single edge or to two adjacent edges vanish in our gauge and are not displayed.}
\label{fig:3typesofdiagrams}
\end{figure}

In the this section we calculate the two-loop contributions to the 
tetragonal light-like Wilson loop in pure Chern Simons theory. 
The results are 
consistent with
the anomalous conformal Ward identity to be discussed
in section \ref{sec:anomalous-ward-identities}.

Expanding the Wilson loop to quartic order, see \eqref{eqn:Wilson-Loop-after-path-ordering}, and 
performing Wick contractions leads to the topologies shown in Figure \ref{fig:3typesofdiagrams}.
We are taking the planar limit and therefore drop all non-planar graphs.
Moreover, all diagrams where one propagator is connected to a single edge or adjacent edges vanish
in our gauge for the same reason as at the one-loop order and are not displayed.

\subsection{Ladder diagrams}
Let us begin by computing diagrams of ladder topology as shown in Figure \ref{fig: rect1331}.
There are two different orientations of this diagram, and it is easy to see that they give the same
contribution. Taking into account this factor of $2$, we have that the contribution of the ladder
diagrams is
\begin{align}
\langle W_4 \rangle^{(2)}_{\text{ladder}}
&= 2 \left( \frac{N}{k} \right)^2 \left(   \frac{\Gamma\left(\frac{d}{2}\right)}{\pi^{\frac{d-2}{2}}} \right)^2  \, I_{\text{ladder}}( x_{13}^2 , x_{24}^2 )\,,
\end{align}
where
\begin{align}
I_{\text{ladder}}(x_{13}^2, x_{24}^2) 
% &=\frac{1}{N}(i)^4  \oint_{{z_i>z_j>z_k>z_l}} \hspace{-45pt} dz_{i,j,k,l}^{\mu,\nu,\rho,\sigma} \langle \tr A_\mu(z_i) A_\nu(z_j) A_\rho(z_k) A_\sigma(z_l) \rangle \\ \nonumber
&=   \int ds_{i,j,k,l} \frac{\epsilon(\dot{z}_i,\dot{z}_l,z_i-z_l)}{[-(z_i-z_l)^2]^{\frac{d}{2}}} \frac{\epsilon(\dot{z}_j,\dot{z}_k,z_j-z_k)}{[-(z_j-z_k)^2]^{\frac{d}{2}}} \,. 
\end{align}
The integral is finite and may be calculated for $d=3$
\begin{align}
I_{\text{ladder}}(x_{13}^2,x_{24}^2) &= \frac{1}{4} \int_0^1 ds_i \int_0^{s_i} ds_j \int_0^1 ds_k \int_0^{s_k} ds_l \frac{ x_{13}^2 x_{24}^2 ( x_{13}^2 +x_{24}^2)}{[x_{13}^2 \bar{s}_i \bar{s}_l+ x_{24}^2 s_i s_l]^{\frac{3}{2}} [x_{13}^2 \bar{s}_j \bar{s}_k+ x_{24}^2 s_j s_k]^{\frac{3}{2}}}  + O(\epsilon) \,.
\end{align}
We computed this integral by first carrying out some of the parameter integrals and then deriving a differential equation for it, which could be solved.
The result is remarkably simple, 
\begin{align} 
I_{\text{ladder}}(x_{13}^2,x_{24}^2) &= \frac{1}{2} \left[ \ln^2 \left(\frac{x_{13}^2}{x_{24}^2}\right) + \pi^2\right] + O(\epsilon) \,.
\end{align}
Including the prefactors and dropping $O(\epsilon)$ terms, the contribution to the Wilson loop is
\begin{align}\label{eqn:result-ladder}
\langle W_4 \rangle^{(2)}_{\text{ladder}}
&=  \left( \frac{N}{k} \right)^2   ~\frac{1}{4} \left[ \ln^2 \left(\frac{x_{13}^2}{x_{24}^2}\right) + \pi^2\right]\,.
\end{align}

\subsection{Vertex diagrams}
\label{sect:vertex-diagram}

The diagrams with one three-gluon vertex shown in Figures \ref{fig:vertexI111} , \ref{fig:vertexI113} and \ref{fig:vertexI123}
are obtained by contracting the cubic term in the expansion of the 
Wilson loop in \eqref{eqn:Wilson-Loop-after-path-ordering} with the interaction term of the 
Lagrangian,
\begin{align}\label{eqn:vertex-diagram}
\langle W_4 \rangle^{(2)}_{\text{vertex}}
%&= \frac{1}{N} \langle (i)^3  \oint_{z_i>z_j>z_k} \hspace{-30pt} dz_{i,j,k}^{\mu,\nu,\rho}\, \tr \left( A_\mu A_\nu A_\rho \right) \left( i \int\, d^dw \mathcal{L}_{\text{int}}(w)\right)\rangle \\ \nonumber
%&= - \frac{1}{N} \frac{k}{4 \pi} \frac{2}{3}  (i)^5 \oint dz_{i,j,k}^{\mu,\nu,\rho}\int d^dw \langle \tr \left( A_\mu A_\nu A_\rho \right) \tr \left( A_\alpha A_\beta A_\gamma (w) \right) \epsilon^{\alpha\beta\gamma}\rangle \\ \nonumber
&=   \left(\frac{N}{k}\right)^2 \frac{i}{2 \pi}  \left( \frac{\Gamma\left(\frac{d}{2}\right)}{\pi^{\frac{d-2}{2}}}\right)^3 \sum_{i>j>k} I_{ijk} \,,
\end{align}
where 
\begin{equation}\label{eqn:Iijk}
I_{ijk}= -\int dz_i^\mu dz_j^\nu dz_k^{\rho}  \epsilon^{\alpha\beta\gamma}\epsilon_{\mu\alpha\sigma}\epsilon_{\nu\beta\lambda}\epsilon_{\rho\gamma\tau}\int d^dw \frac{(w-z_i)^{\sigma}(w-z_j)^{\lambda}(w-z_k)^{\tau}}{|w-z_i|^d|w-z_j|^d|w-z_k|^d}\,,
\end{equation}
and $|z_i|=(-z_i^2)^{\frac{1}{2}}$.
Here the indices of $I_{ijk}$ refer to the edges of the Wilson loop that the propagators attach to.
The expression can be shown to be antisymmetric under the exchange of any two indices,
and therefore the only non-vanishing contributions are the ones for $i\neq j\neq k$. 
As a consequence, topologies  \ref{fig:vertexI111} and \ref{fig:vertexI113} can be discarded.

Specialising to the tetragon, we have four contributions which are symmetric under $x_{13}^2 \leftrightarrow  x_{24}^2$ 
and thus it is sufficient to compute one of them
\begin{align}\label{eqn:vertex-Iijk}
I_{321}&%=-I_{123}
= \int d^dw \int_0^1 ds_{1,2,3} \frac{\epsilon(p_2,p_3,w)\epsilon(p_2,p_1,w)}{|w|^{d}|w-z_{12}|^{d}|w-z_{32}|^{d}} \\ \nonumber
%&\stackrel{\eqref{app:Vertex}}{=} 
&=
\frac{i \pi^{\frac{d}{2}}}{8}\frac{\Gamma(d-1)}{\Gamma\left(\frac{d}{2} \right)^3} x_{13}^2  x_{24}^2  \int_0^1 d^3 s_{1,2,3} d^3\beta_{1,2,3} \left(\beta_1\beta_2\beta_3 \right)^{\frac{d-2}{2}}\delta \left(\sum_{i=1}^{3} \beta_i-1\right)  \times \nonumber \\
& \qquad\qquad\qquad \qquad\qquad  \times \left( \frac{1}{\Delta^{d-1}}- 2 \frac{(d-1)}{\Delta^d} \beta_1 \beta_3 \bar{s}_1 s_3 (x_{13}^2+x_{24}^2)\right) \,, \nonumber
\end{align}
where the second and third line is obtained by introducing Feynman parameters in the standard way and  integrating over $w$. More details
may be found in Appendix \ref{app:Vertex}. $\Delta$ is given by
\begin{align}\label{delta-vertex}
\Delta &= -\beta_{1} \beta_{2} z_{12}^2 - \beta_{2} \beta_{3} z_{23}^2 - \beta_{1} \beta_{3} z_{13}^2  \nonumber \\
&= -x_{13}^2 \beta_1 \bar{s}_1 \left(\beta_3 \bar{s}_3+ \beta_2 s_2 \right) - x_{24}^2 \beta_3 s_3 \left(\beta_2 \bar{s}_2 + \beta_1 s_1  \right)
\end{align}

\begin{figure}[t]
\centering
	\includegraphics[width=.15 \textwidth]{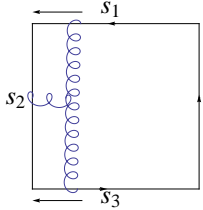}
\caption{The divergence in the vertex diagram arises from the integration region where 
$s_1 \rightarrow 1, {s}_3 \rightarrow 0$
(and $\beta_{1} \rightarrow 0, \beta_{3} \rightarrow 0 $), see equation (\ref{delta-vertex}).
%$s_1 , \bar{s}_3\rightarrow 0$ and $s_1, \bar{s}_3\rightarrow 1$.
}
\label{fig:divergence-vertex}
\end{figure}
One might naively think that this diagram should give a finite answer due to the antisymmetry of the $\epsilon$ tensors.
The result would indeed be finite in the case of smooth contours \cite{Guadagnini:1989am} or contours with a single cusp. 
However, the presence of two cusps gives rise to a region in the integration space of Feynman parameters
where the first summand  in the third line of \eqref{eqn:vertex-Iijk} induces a divergent contribution.
The relevant region of Feynman parameters is 
%$s_1, \bar{s}_3\rightarrow 0$ and $s_1, \bar{s}_3\rightarrow 1$
$s_1 \rightarrow 1, {s}_3\rightarrow 0$
(and $\beta_{1} \rightarrow 0, \beta_{3} \rightarrow 0$), see equation \eqref{delta-vertex},
and is illustrated in Figure \ref{fig:divergence-vertex}. Due to the presence of three independent vectors
$p_{1}^{\mu}, p_{2}^{\mu}$ and $p_{3}^{\mu}$ the $\epsilon$ tensors do not suppress this region.
We find that this term produces a $1/\epsilon$ pole in dimensional reduction.
The second summand in the third line of \eqref{eqn:vertex-Iijk} is finite. 

We separated the divergent and finite pieces using Mellin-Barnes techniques. The details
can be found in Appendix \ref{app:Vertex}. We have not computed the coefficients of the $\epsilon^{-1}$ and
$\epsilon^{0}$ terms analytically, but we have good numerical evidence that they give the following result:
\begin{equation}
I_{321}= \frac{i \pi^{\frac{d}{2}+1}}{8}\frac{\Gamma(d-1)}{\Gamma\left(\frac{d}{2} \right)^3}  \left[ 2 \ln(2)\frac{ (-x_{13}^2)^{2\epsilon}+(-x_{24}^2 )^{2\epsilon}}{\epsilon}  + \ln^2 \left(\frac{x_{13}^2}{x_{24}^2}\right) + a_6 + \mathcal{O}(\epsilon) \right]
\end{equation}
where $a_6 =  8.354242685 \pm 2 \cdot 10^{{-9}} $, see  \eqref{eqn:a6}. 
Taking into account all prefactors 
and restoring the regularisation scale, $k \rightarrow \mu^{-2\epsilon} k$, we can write
%\footnote{We have rewritten  $\frac{\Gamma(2-2 \epsilon)}{\epsilon}=
%\frac{(e^{\gamma_E})^{2\epsilon}}{\epsilon}-2+\mathcal{O}(\epsilon)$, 
% which appears in the prefactors and $c_2$.} 
the result, up to terms of order $\epsilon$,  as
\begin{align}\label{eqn:result-vertex}
\langle W_4 \rangle^{(2)}_{\text{vertex}}&= -\left( \frac{N}{k}\right)^2  \left[ \frac{\ln(2)}{4}  \sum_{i=1}^{4} \frac{(-x_{i,i+2}^2 \, \mu^2 \pi e^{\gamma_E})^{2\epsilon}}{\epsilon}+ \frac{1}{4} \ln^2\left( \frac{x_{13}^2}{x_{24}^2} \right) + \frac{1}{4} a_6- 2 \ln(2)  \right]\,.
\end{align}

\subsection{Gauge field and ghost loops}
It is well known \cite{Chen:1992ee} that in the dimensional reduction (DRED) scheme
%i.e. performing algebraic operations with the epsilon-tensors in strictly 3 dimensions, 
the gauge field loop diagrams shown in Figure \ref{fig:gluon-loop} exactly cancel against the ghost loop diagrams 
shown in Figure \ref{fig:ghost-loop} :
%\footnote{Using the DREG scheme, the contributions cancel up to an order $\epsilon$ term}
\begin{align}\label{eqn:gluon-ghost-loops}
\langle W_4 \rangle^{(2)}_{\text{gluon loop}} = -  \langle W_4 \rangle^{(2)}_{\text{ghost loop}} \,.
\end{align}
Details of this cancellation can be found in appendix \ref{app:gluon-and-ghost-loops}.

\subsection{Result for the two-loop tetragon in CS theory}
\label{sect:CS-2loop-results}
Summing up the results \eqref{eqn:result-ladder}, \eqref{eqn:result-vertex} and \eqref{eqn:gluon-ghost-loops} for the tetragon, 
interestingly, the  $\ln^2(x_{13}^2/ x_{24}^2)$
terms in the two-gauge-field diagram and the vertex diagram 
exactly cancel and we obtain
\begin{align}\label{eqn:result-wilson-loop-two-loop}
\langle W_4 \rangle^{(2)} &=  -\left(\frac{N}{k}\right)^2  \frac{1}{4}\left[\ln(2) \sum_{i=1}^4\frac{(-x_{i,i+2}^2 \, \tilde \mu^2 )^{2\epsilon}}{\epsilon} + a_6 - 8 \ln(2) - \pi^2 \right]\,.
\end{align}
where $\tilde \mu^2 = \mu^2 \pi e^{\gamma_E}$ and we recall that $a_6 =  8.354242685 \pm 
2 \cdot 10^{{-9}} $.
As we will see in the next section, the cancellation observed here that led to the finite part of \eqref{eqn:result-wilson-loop-two-loop} being
a constant is in fact a consequence of the (broken) conformal symmetry of the Wilson loops under consideration.

\section{Anomalous conformal Ward  identities}\label{sec:anomalous-ward-identities}
The structure of the above results can be understood from conformal symmetry,
by deriving anomalous conformal Ward identities for the Wilson loops.
Here we follow very closely reference \cite{Drummond:2007au} .

We would like to use the specific properties of 
Wilson loops with light-like contours $\mathcal{C}$ under conformal transformations.  
The key point is that such contours are stable under conformal transformations, 
i.e. the deformed contour $\mathcal{C}^\prime$ is also made of $n$ light-like segments. 
This can be seen as follows.
The cusp points $x_i$ form a contour with light-like edges, i.e. $x_{i,i+1}^2=0$. It is obvious that the light-likeness conditions are preserved by translations, rotations, and dilatations.
Special conformal transformations 
are equivalent to an inversion $x_\mu \rightarrow x_\mu/x^2$ followed 
by a translation and another inversion. 
Thus it remains to investigate the transformation under inversions. 
Since under the latter 
$x_{ij}^2 \rightarrow %x_{ij}^{\prime\, 2}=
{x_{ij}^2}/{(x_i^2 x_j^2)}$, it is clear that 
the light-likeness of the contour is preserved by all conformal transformations. 

If the Wilson loop $\langle W_n \rangle $ were well defined in $d=3$ dimensional Minkowski space 
it would enjoy the conformal invariance of the underlying gauge theory 
and 
we would conclude that
$ \langle W(\mathcal{C}) \rangle = \langle W(\mathcal{C}^\prime) \rangle$.
This is indeed the case at one loop order, see section \ref{sec:one-loop}.
However, as we have seen in section \ref{sec:two-loop}, starting from two loops, divergences force 
us to introduce a regularisation and calculate in $d=3-2\epsilon$ dimensions, 
thereby breaking the conformal invariance of the action.  The latter leads to an anomalous term in the
conformal Ward identities for the Wilson loops, as we will see presently. 

The expectation value of the Wilson loop can be written as a functional integral
\begin{equation}\label{eqn:path-integ}
\langle W_n \rangle = \int \mathcal{D}A\, e^{iS_{\epsilon}} \tr \left[\mathcal{P} \exp\left(i \oint_{C_n} dz^\mu 
A_\mu(z)\right) \right]\,, \qquad S_\epsilon = \frac{1}{ \mu^{2\epsilon}} \int d^dx\, \mathcal{L}_{\text{CS}}(x)
\end{equation}
where $\mu$ is the regularisation scale that keeps the action  dimensionless in $d=3-2\epsilon$.
The path-ordered exponential is invariant
under dilatations and the Lagrangian is covariant with weight $\Delta_\mathcal{L}=3$, whereas the 
measure $d^dx$ does not match this weight for $d=3-2\epsilon$. This results in a non-vanishing 
variation of the action with respect to dilatations and special conformal transformations. The conformal 
Ward identities can be derived by acting on both sides\footnote{Note that the left-hand side of 
\eqref{eqn:path-integ}
is a function 
of the cusp points, whereas its right-hand side contains all fields of the Lagrangian.} of \eqref{eqn:path-integ} 
with the generators of conformal transformations, see \cite{Drummond:2007au,Sarkar:1974xh,Braun:2003rp}. 
This leads to the Ward identities
\begin{align}\label{eqn:dilatation-Wardidentitiy}
\mathbb{D}\,  \langle W_n  \rangle &= -\frac{2 i \epsilon}{ \mu^{2\epsilon}} \int d^dx \langle \mathcal{L}(x) W_n \rangle \,, \\ \label{eqn:special-conformal-Wardidentitiy}
\mathbb{K}^\nu  \langle W_n  \rangle &= -\frac{4 i \epsilon}{\mu^{2\epsilon}} \int d^dx\, x^\nu \langle \mathcal{L}(x) W_n \rangle \,,
\end{align}
for {dilatations} and {special conformal} transformations. Here the operators on the left hand sides act in
the canonical way on the coordinates of the cusp points,
\begin{align}\label{eqn:diff-equ-ward-id}
\mathbb{D}    &= \sum_i (x_i \cdot \partial_i)  \,,\qquad  
\mathbb{K}^\nu  = \sum_i \left(2 x_i^\nu (x_i\cdot \partial_i)- x_i^2 \partial_i^\nu   \right) \,.
\end{align}
We emphasise that thanks to the factor of $\epsilon$ on the r.h.s. of \eqref{eqn:dilatation-Wardidentitiy} and \eqref{eqn:special-conformal-Wardidentitiy} 
it is sufficient to know the divergent part of the integrals appearing on the r.h.s. of those equations in order to obtain information about
the finite part of $\langle W_n \rangle $.

The dimensionally regularised  Wilson loop $\langle W_n \rangle$ is a dimensionless scalar function of 
the cusp points $x_i^\nu$, which appear paired with the regularisation scale as $x_{ij}^2 \mu^{2}$. 
As a consequence $\langle W_n \rangle$ satisfies
\begin{equation}\label{eqn:wilson-loop-regularisation}
\left( \sum_{i=1}^n (x_i \cdot \partial_i) - \mu\, \frac{\partial}{\partial\mu} \right) \langle W_n \rangle =0 \,.
\end{equation}
This provides a consistency condition for the right hand side of \eqref{eqn:dilatation-Wardidentitiy}.

\subsection{One-loop insertions}
\begin{figure}[h]
\centering
	\includegraphics[width=0.15 \textwidth]{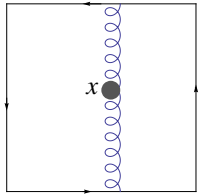}
	\caption{Lagrangian insertion contributing to the Ward identities at one loop. }
	\label{fig:lagrangian-insertion-1-loop}
\end{figure}

At order  $N/k$ we have a contribution from the contraction of the kinetic part of the Lagrangian 
insertion with the second order expansion of the Wilson loop operator, shown in fig. \ref{fig:lagrangian-insertion-1-loop},
\begin{align}
\langle \mathcal{L}(x) W_n \rangle^{(1)}
= \langle \mathcal{L}_{\text{kin}}(x) W_n\rangle^{(1)} = \frac{(i)^2}{N}\int_{z_i>z_j} \hspace{-20pt} dz_{i,j}^{\mu,\nu} \epsilon^{\alpha\beta\gamma} \langle \tr ( A_\alpha \partial_\beta A_\gamma)(x) \tr (A_\mu A_\nu) \rangle^{(1)} \,.
\end{align}
The direct calculation of the right hand sides of
\eqref{eqn:dilatation-Wardidentitiy} and \eqref{eqn:special-conformal-Wardidentitiy}, 
yields a vanishing result as $\epsilon \rightarrow 0$. 
%For the dilatation Ward identity this is easy to see, since the integration 
%of 
%\begin{equation}
%\langle A_\mu(z_i) A_\alpha(x) \rangle \partial_\beta^{(x)} \langle A_\nu(z_j) A_\gamma(x) \rangle 
%\end{equation}
%over $d^dx$ 
%and contraction with the Levi-Civita symbols effectively yields\footnote{The same integral appears in \eqref{eqn:integration-yields-prop}. } a gluon propagator of type $\langle A_\mu(z_i) A_\nu(z_j) \rangle$. Thus, the dilatation Ward identity is equivalent to the one-loop diagrams which are all finite. Furthermore, one finds that there are no contributions to the special conformal Ward identity either.\\
Thus we have 
\begin{align}\label{eqn:ward-oneloop}
\mathbb{D}\langle W_n  \rangle^{(1)} =  O(\epsilon)\,, \qquad {\rm and} \qquad \mathbb{K}^\nu \langle W_n  \rangle^{(1)}= O(\epsilon)\,,
\end{align} 
in other words the conformal symmetry is unbroken for $\epsilon = 0$.
As a consequence, the expectation value of the Wilson loop
is constrained to be a function of conformally invariant variables.
Starting from the Lorentz invariants $x_{ij}^2$ the most general
conformal invariants are the cross-ratios
\begin{equation}\label{eqn:cross-ratios}
u_{ijkl} := \frac{x_{ij}^2 x_{kl}^2}{x_{il}^2 x_{jk}^2}\,.
\end{equation}
In our case where neighbouring points are light-like separated, $x_{i,i+1}^2=0$,
non-vanishing cross-ratios can only be written down starting from $n=6$.
The special conformal Ward identities \eqref{eqn:ward-oneloop} then imply that $\langle W_n \rangle^{(1)}$
is given by a function of conformal cross-ratios,
\begin{equation}\label{eqn:sol-sc-WI-one-loop}
\langle W_n \rangle^{(1)} = g_{n}\left( u_{ijkl}\right)\,, \qquad \langle W_4 \rangle^{(1)} = {\it const}. 
\end{equation}
Since there are no non-vanishing conformal cross-ratios at four points, $ \langle W_4 \rangle^{(1)} $ must be a constant.

Let us now compare against the results of our one-loop computation of section \ref{sec:one-loop}.
There, the constant on the r.h.s. of the second equation in \eqref{eqn:sol-sc-WI-one-loop} was 
found to be zero for the tetragon. 
Moreover, analytical investigations of certain symmetric contours and numerical investigations 
for non-symmetric contours show that $g_{6}(u_{ijkl})$ is zero for the hexagon. 
As mentioned before, we 
expect that the result remains true for higher polygons, i.e. $g_{n} = 0$.

\subsection{Two-loop insertions}
\begin{figure}[t]
\centering
\begin{tabular}{ c c c c c c c}
% First ROW %%%%%%%%%%%%%
%\raisebox{35pt}{ $\langle \mathcal{L}(x)  W_n \rangle$ =} 
~& 
\subfloat[]{\label{fig:kinetic-insertion}
\includegraphics[width=.18 \textwidth]{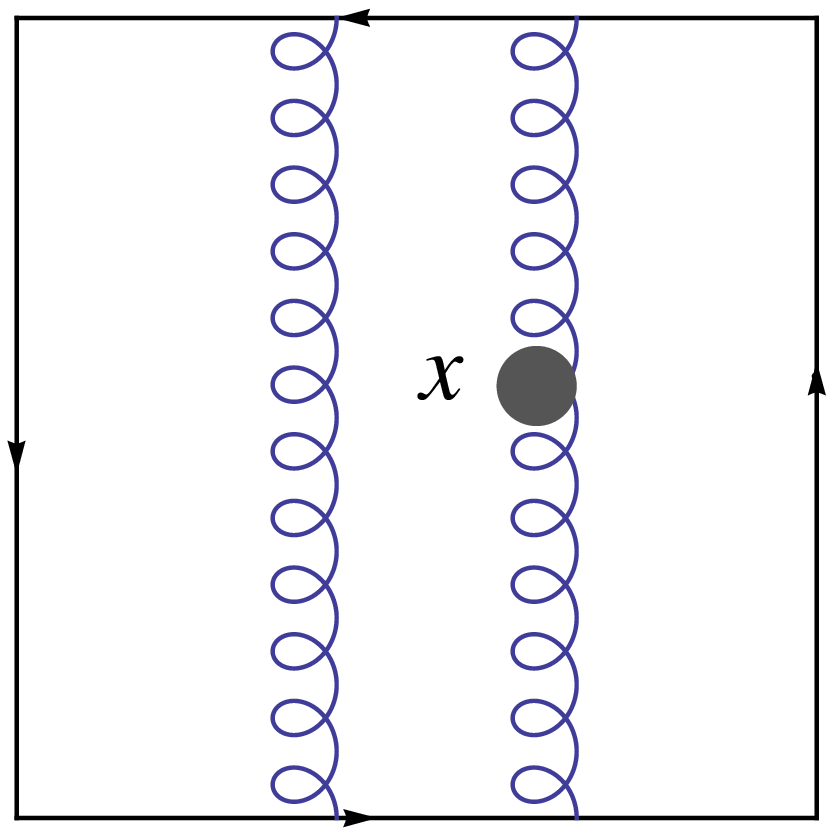}} & % \raisebox{35pt}{ + } 
&
\subfloat[]{\label{fig:vertex-insertion}
\includegraphics[width=.18 \textwidth]{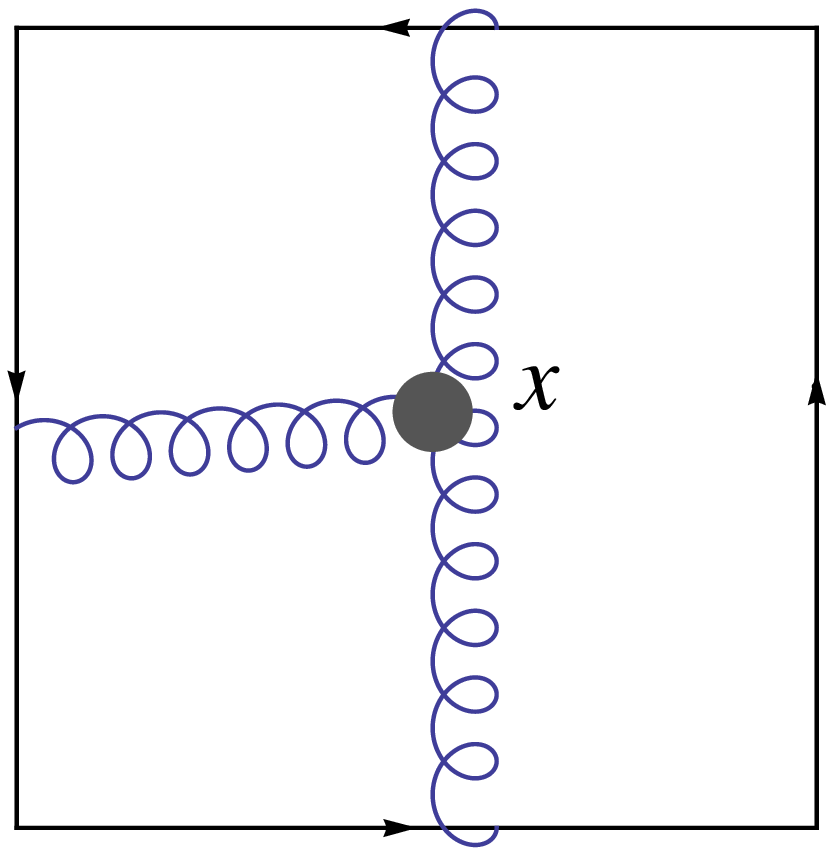}} & 
%\raisebox{35pt}{ + }
&
\subfloat[]{\label{fig:kinetic-insertion-in-vertex}
\includegraphics[width=.18 \textwidth]{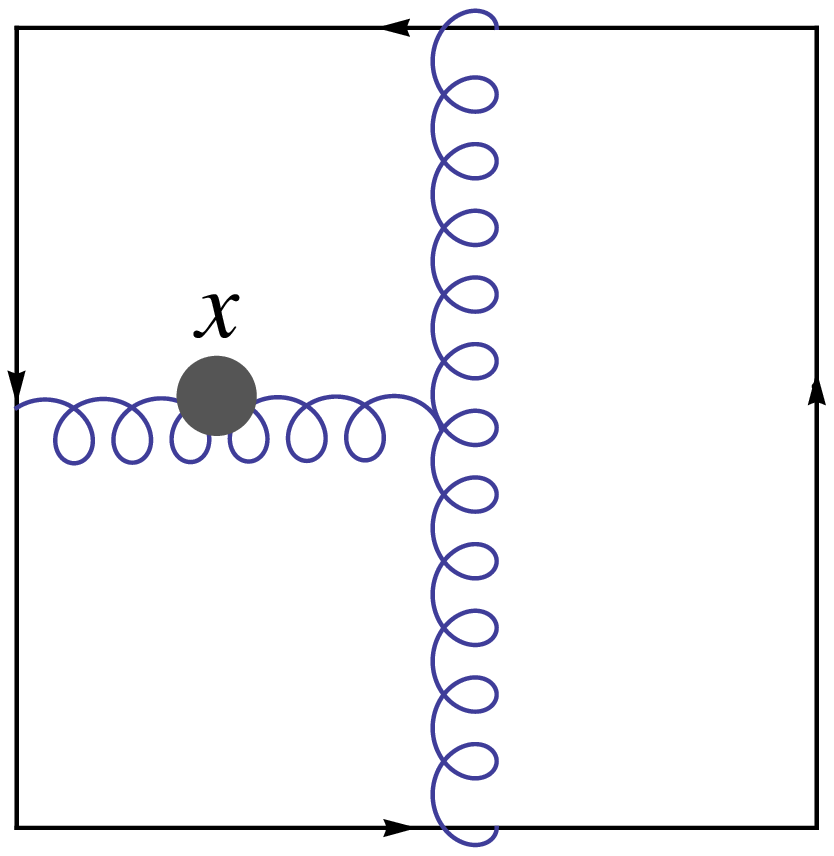}} &
%\raisebox{35pt}{ + }   
\\
% Second ROW %%%%%%%%%%%%
~& \subfloat[]{\label{fig:gluon-loop-vertex-insertion}
\includegraphics[width=.185 \textwidth]{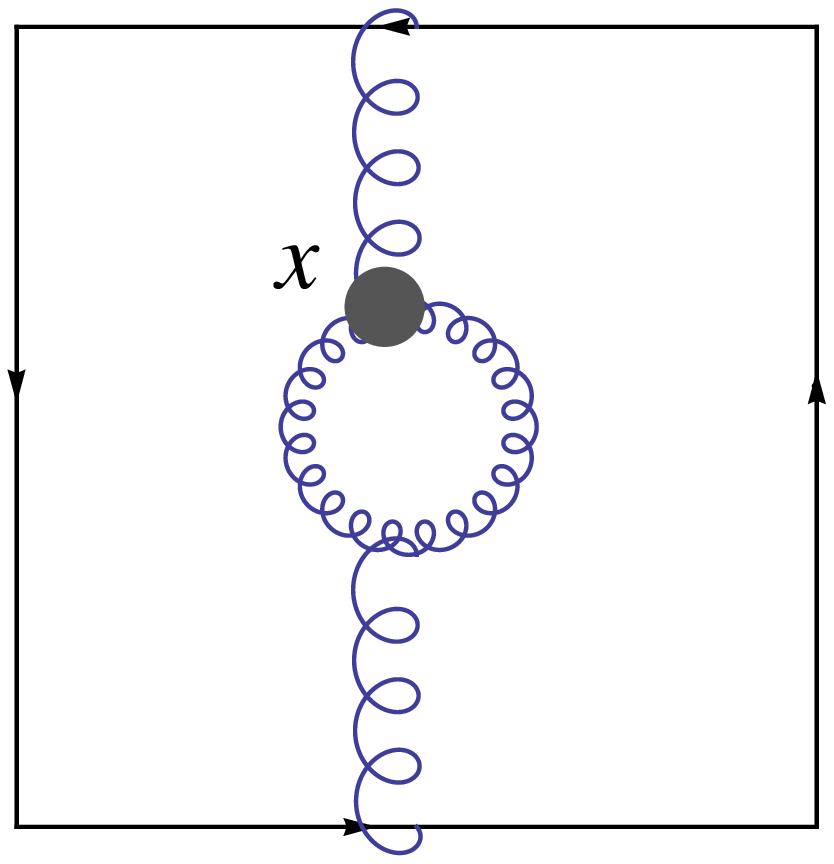}} & 
%\raisebox{35pt}{ + } 
&
\subfloat[]{\label{fig:gluon-loop-propagator-insertion}
\includegraphics[width=.185 \textwidth]{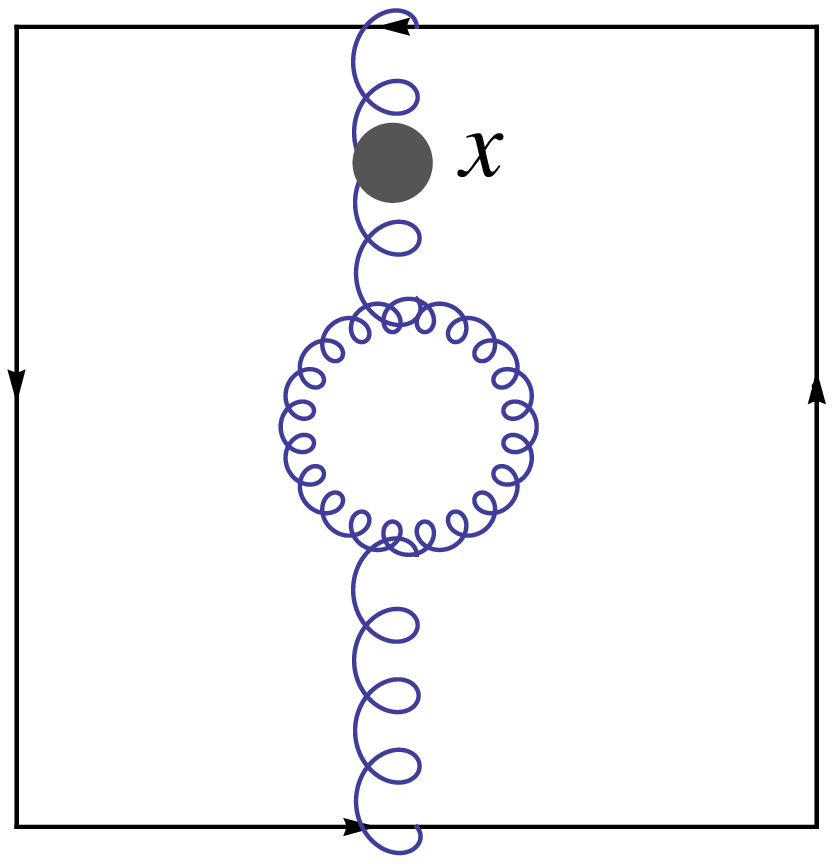}} & 
%\raisebox{35pt}{ + } 
&
\subfloat[]{\label{fig:gluon-loop-inloop-insertion}
\includegraphics[width=.185 \textwidth]{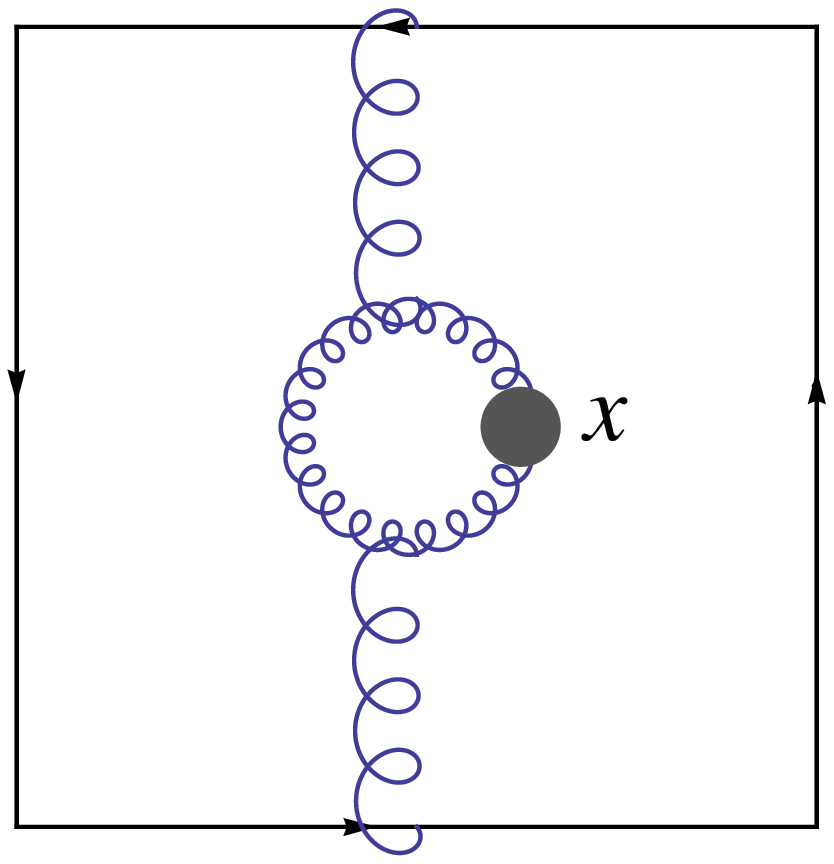} } &
%\raisebox{35pt}{ + }  
\\
% Third ROW %%%%%%%%%%%%%
~& \subfloat[]{\label{fig:ghost-loop-vertex-insertion}
\includegraphics[width=.185 \textwidth]{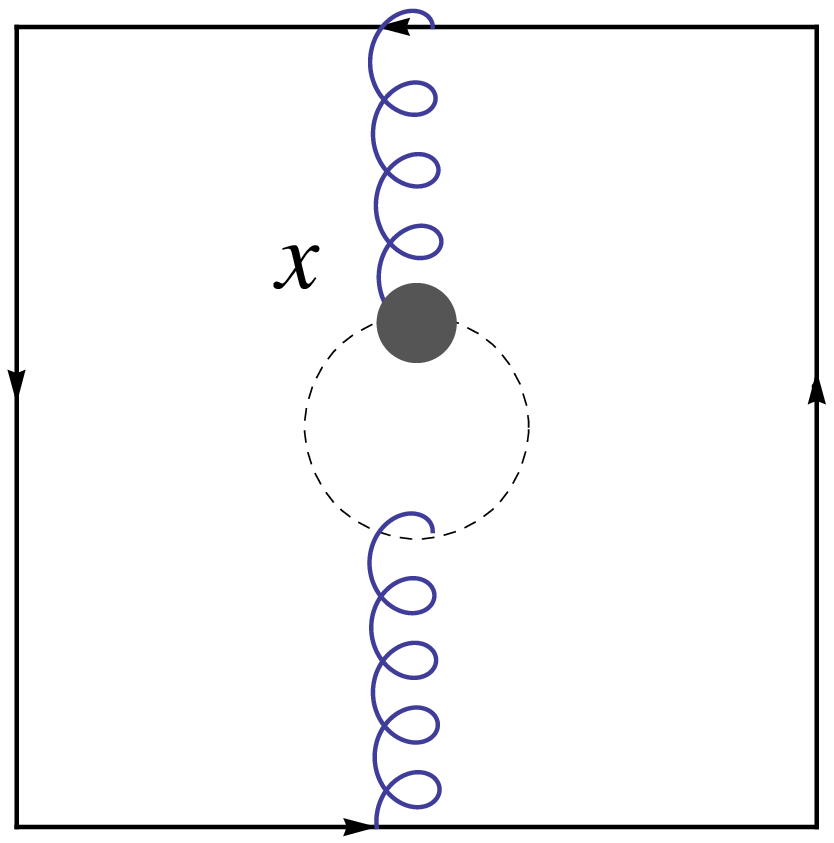}} & 
%\raisebox{35pt}{ + } 
&
\subfloat[]{\label{fig:ghost-loop-propagator-insertion}
\includegraphics[width=.185 \textwidth]{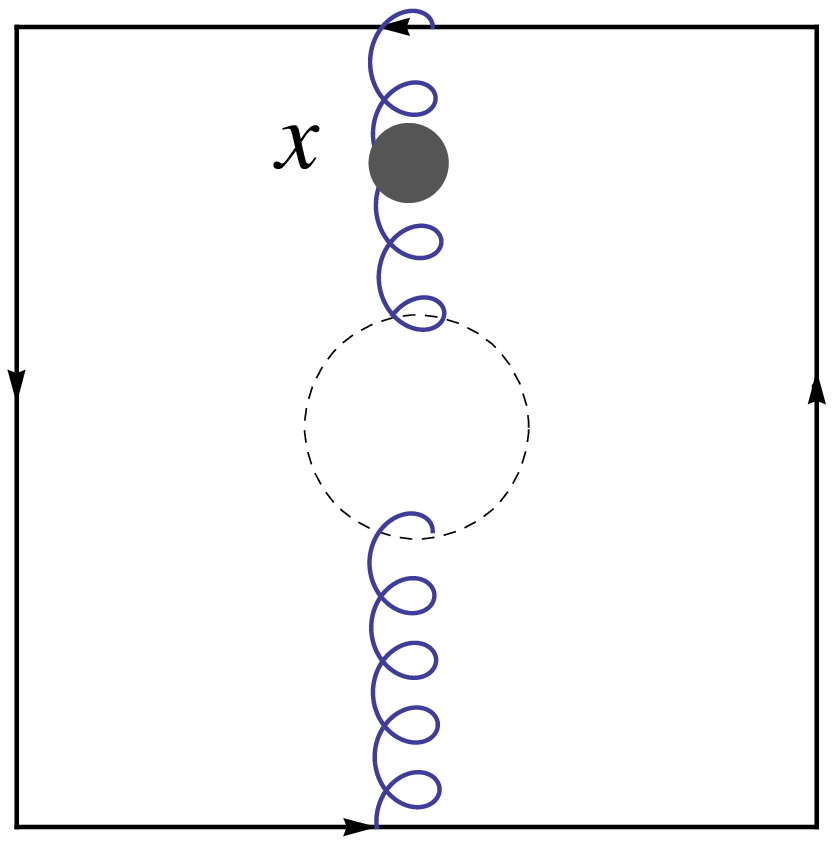}} & 
%\raisebox{35pt}{ + } 
&
\subfloat[]{\label{fig:ghost-loop-inloop-insertion}
\includegraphics[width=.185 \textwidth]{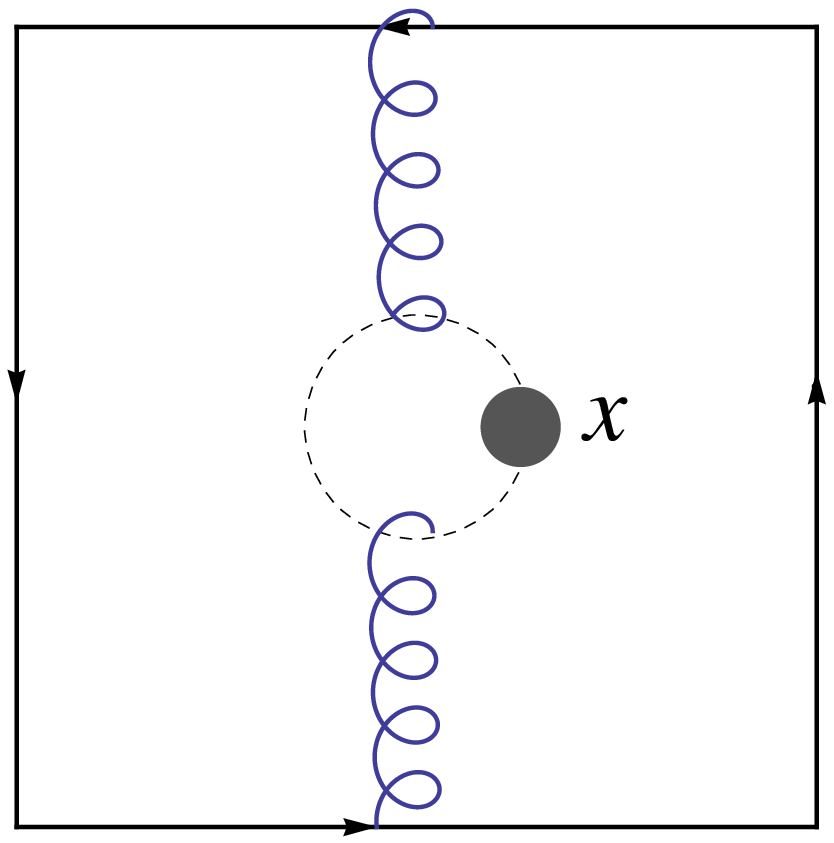}} & ~
\end{tabular}
\caption{Two-loop contributions to the anomalous conformal Ward identity. 
The grey point in the pictures denotes the point $x$ of the Lagrangian insertion 
in $\langle \mathcal{L}(x) W_n \rangle$. 
The diagrams of the second and third line cancel pairwise.
}
\label{fig:2-loop-diagrams-anomalous-conformal-ward-identity}
\end{figure}

At two loops there are several diagrams that contribute to the insertion of the Lagrangian into the
Wilson loop,
$ \langle \mathcal{L}(x) W_n \rangle $,
that correspond to the kinetic term, the gauge field vertex, the ghost kinetic term and the ghost vertex in $\mathcal{L}(x)$. 
Those diagrams are shown in Figure \ref{fig:2-loop-diagrams-anomalous-conformal-ward-identity}. 
We do not display diagrams that vanish for kinematical reasons as at one-loop level.

Just as at one-loop level, only diagrams giving rise to divergent integrals will
contribute to the anomalous Ward identities.

\subsubsection{Insertion into the ladder diagram}
%First, we can contract the kinetic term of the insertion with the quartic order expansion of the Wilson loop, this corresponds to fig. \ref{fig:kinetic-insertion} 
%\begin{align}\label{eqn:insertion-kinetic-term}
%  \langle \mathcal{L}(x) W_4 \rangle^{(2)}_{\text{(a)}} = \langle \mathcal{L}_{\text{kin}}(x) \frac{1}{N}(i)^4 \int dz_{i,j,k,l}^{\mu,\nu,\rho,\sigma}\, \tr \left( A_\mu A_\nu A_\rho A_\sigma \right)  \rangle \,.
% \end{align}
Let us consider the insertion of the kinetic term of the action into the ladder diagram as shown in Figure \ref{fig:kinetic-insertion}.
For the dilatation Ward identity these are exactly the two-gluon diagrams calculated above, which are finite. 
For the special conformal Ward identity the integration is slightly more complicated, but the finiteness is easy to check for all contributions. 
%Thus we have
%\begin{align}
%\mathbb{D}:&\qquad \epsilon \int d^dx  \langle \mathcal{L}(x) W_4 \rangle^{(2)}_{\text{(a)}} = \mathcal{O}(\epsilon)\,, \\ \nonumber
%\mathbb{K}^\nu:& \qquad\epsilon \int d^dx\, x^\nu  \langle \mathcal{L}(x) W_4 \rangle^{(2)}_{\text{(a)}} = \mathcal{O}(\epsilon)\,.
%\end{align}
Thus, this diagram does not contribute to the anomalous Ward identities.

\subsubsection{Insertion of the interaction term}
Next, we can contract the cubic order expansion of the Wilson loop 
with the vertex term of the Lagrangian insertion, 
as shown in Figure \ref{fig:vertex-insertion},
\begin{align}
\langle \mathcal{L}(x) W_4 \rangle^{(2)}_{\text{(b)}} 
&= \langle \mathcal{L}_{\text{int}}(x) \frac{1}{N}(i)^3 \int d^3z_{i,j,k}^{\mu,\nu\rho}\, \tr \left( A_\mu A_\nu A_\rho \right)  \rangle \,.
\end{align}
For the dilatation Ward identity, we trivially  have
\begin{align}\label{eqn:result-insertion-interaction-term}
\int d^dx  \langle \mathcal{L}(x) W_4 \rangle^{(2)}_{\text{(b)}} 
&= \frac{1}{i} \langle W_4 \rangle^{(2)}_\text{vertex} = 
i \left( \frac{N}{k}\right)^2   \frac{\ln(2)}{\epsilon}  
+ O(\epsilon^0 )\,,
\end{align}
which is just the vertex diagram that was calculated in \eqref{eqn:vertex-diagram}, up to a factor of $i$. 

The contribution to the special conformal Ward identity is more complicated. 
We have
\begin{align}\label{eqn:special-conformal-wi-vertex-insertion}
\int d^dx\,  x^\nu \langle \mathcal{L}(x) W_4 \rangle^{(2)}_{\text{(b)}} 
&= \left( \frac{N}{k} \right)^2 \frac{1}{2\pi}   \left( \frac{\Gamma\left(\frac{d}{2}\right)}{\pi^{\frac{d-2}{2}}}\right)^3 \sum_{i>j>k} I^\prime_{ijk} 
\end{align}
with
\begin{align}\label{eqn:I321prime}
I^\prime_{321} =& \int_0^1 ds_{1,2,3} \int d^dx (x + z_2)^\nu   \frac{\epsilon(p_2,p_3,x)\epsilon(p_2,p_1,x)}{|x|^{d}|x-z_{12}|^{d}|x-z_{32}|^{d}}
\\
=&  \frac{2 \pi i \ln (2)}{\epsilon} (x_2 + x_3)^\nu + \mathcal{O}(\epsilon^0)\,,
\end{align}
where the coefficient $\ln(2)$ was computed numerically to $10$ relevant digits. The reason the pole arises was discussed in section \ref{sect:vertex-diagram}.

Details of this calculation can be found in appendix \ref{app:conf-wi-insertion-interaction-term}. Summing up all four contributions $I^\prime_{321}, I^\prime_{421}, I^\prime_{432}, I^\prime_{431}$ we arrive at
\begin{align}\label{eqn:special-conformal-wi-vertex-insertion}
\int d^dx\,  x^\nu \langle \mathcal{L}(x) W_4 \rangle^{(2)}_{\text{(b)}} 
&=  \frac{i}{\epsilon}  \left( \frac{N}{k} \right)^2  \frac{\ln(2)}{4} \, \sum_{i=1}^{4} x_i^\nu \,.
\end{align}

\subsubsection{Insertion of the kinetic term into the vertex diagram}
Furthermore, we can contract one gauge field of the kinetic term of the insertion with the 
Wilson loop and the other one with the 3-gauge-field vertex, leading to a diagram of the type 
displayed in Figure \ref{fig:kinetic-insertion-in-vertex},
\begin{align}\label{eqn:kinetic-insertion-in-vertex}
\langle \mathcal{L}(x) W_4 \rangle^{(2)}_{\text{(c)}} 
&=  \langle \mathcal{L}_{\text{kin}}(x) \frac{1}{N} \oint_{z_i > z_j > z_k} 
\hspace{-35pt} dz_{i,j,k}^{\mu,\nu,\rho}\, \tr \left( A_\mu A_\nu A_\rho \right) 
\left( i\int d^dw \mathcal{L}_{\text{int}}  \right)  \rangle \\ \nonumber
=& \left(\frac{k}{4 \pi}\right)^2   (i)^3  \frac{2}{3}\frac{1}{N} \int d^dx \int d^dw \oint dz_{i,j,k}^{\mu,\nu,\rho}\\ \nonumber
&\phantom{=}\qquad \epsilon^{\alpha\beta\gamma}\epsilon^{\delta\sigma\tau} \langle \tr \left(A_\alpha \partial_\beta A_\gamma \right)(x) \tr \left(A_\mu A_\nu A_\rho \right) \tr \left(A_\delta A_\sigma A_\tau \right)(w)  \rangle 
\end{align}
Let us Wick-contract
the kinetic term with $A_\nu(z_j)$ (the two other contractions are discussed below.)
We obtain
\begin{align}
\left( \frac{N}{k}\right)^2 \frac{i}{8 \pi^2}  \left( \frac{\Gamma\left(\frac{d}{2}\right)}{\pi^{\frac{d-2}{2}}}\right)^4 \int d^dw  \oint dz_{i,j,k}^{\mu\nu\rho} \epsilon^{\delta\sigma\tau} I_{\nu\sigma} G_{\mu\tau}(z_i-w) G_{\rho\delta}(z_k-w) \nonumber \,,
\end{align}
where $G_{\mu\nu}= \epsilon_{\mu\nu\rho}{(x-y)^\rho}/{\left(-(x-y)^2\right)^{\frac{d}{2}}}$ 
and where
\begin{align}
I_{\nu\sigma}(x- z_j,x-w) &=\epsilon^{\alpha\beta\gamma}\left[ G_{\alpha\nu}(x-z_j) \partial_\beta^{(x)} G_{\gamma\sigma}(x-w) + G_{\alpha\sigma}(x-w) \partial_\beta^{(x)} G_{\gamma\nu}(x-z_j) \right] \,.
\end{align} 
The only dependence on the insertion point $x$ is in $I_{\nu\sigma}$. For the dilatation Ward 
identity the integral $\int d^dx\, I_{\nu\sigma}$ can easily be computed (for details see Appendix \ref{app:conf-wi-insertion-kinetic-term-in-vertex})  
and effectively gives a propagator such that we have
\begin{align}\label{eqn:result-insertion-kinetic-term-vertex}
\int d^dx\, \langle \mathcal{L}(x) W_4 \rangle^{(2)}_{\text{(c)}} 
&= %- 3 i \left( \frac{N}{k}\right)^2 \frac{1}{2 \pi}  \left( \frac{\Gamma\left(\frac{d}{2}\right)}{\pi^{\frac{d-2}{2}}}\right)^3 \sum_{i>j>k} I_{ijk}=
3 i \langle W_4 \rangle^\text{vertex}
=- 3 i \left( \frac{N}{k}\right)^2   \frac{\ln(2)}{\epsilon}   + O(\epsilon^0 )\,, 
\end{align}
where a factor of 3 was included since the insertion can be in any of the 3 propagators of the vertex diagram and thus 
we get three times the same contribution.

The contribution to the special conformal Ward identity is more complicated, 
since the integration $\int d^dx\, x^\nu I_{\nu\sigma}$ does not just yield a propagator. 
Performing the calculation, we find 
\begin{align}\label{eqn:special-conformal-wi-vertex-insertion-in-propagator}
\int d^dx  \,x^\nu\langle \mathcal{L}(x) W_4 \rangle^{(2)}_{\text{(c)}} 
&= - \frac{i}{\epsilon}  \left( \frac{N}{k} \right)^2 \frac{3}{4}   \ln(2) \, \sum_{i=1}^{4} x_i^\nu + O(\epsilon^0 ) \,,
\end{align}
where the coefficient $\ln(2)$ was computed numerically (details can be found in Appendix \ref{app:conf-wi-insertion-kinetic-term-in-vertex}).

\subsubsection{Insertions with gauge field and ghost loops}
The gauge-field-ghost-insertions in Figure \ref{fig:2-loop-diagrams-anomalous-conformal-ward-identity} cancel pairwise. 
For the dilatation Ward identity the insertions of the three gauge-field and the gauge-field-ghost vertices as shown in diagrams \ref{fig:gluon-loop-vertex-insertion}, \ref{fig:ghost-loop-vertex-insertion} are identical to the gauge field and ghost loop diagrams  \eqref{eqn:gluon-ghost-loops} and thus cancel. It is not necessary to perform the integration over the insertion point to see how the cancellation occurs and thus the contributions to the special conformal Ward identity cancel as well (for details on the cancellation see \ref{app:gluon-and-ghost-loops}).

Insertions of the kinetic term into the gauge field propagator as shown in diagrams \ref{fig:gluon-loop-propagator-insertion}, \ref{fig:ghost-loop-propagator-insertion} cancel as well, since the insertion is the same for both diagrams and thus the algebraic relations responsible for the cancellation in \eqref{eqn:gluon-ghost-loops} remain unchanged. 

Inserting gauge field respectively ghost kinetic terms into the propagators inside the loop  
in diagrams \ref{fig:gluon-loop-inloop-insertion}, \ref{fig:ghost-loop-inloop-insertion} 
produce slightly more complicated expressions.  Nevertheless they cancel as well as can 
be seen in a straightforward calculation. For the dilatation Ward identity the 
cancellation can be seen in an even simpler way by noticing that the integration 
over the insertion point $x$ effectively yields a gauge field respectively ghost propagator. 
Thus the diagrams are identical to the ones in \eqref{eqn:gluon-ghost-loops} and cancel.

\subsection{Anomalous Ward identities and generalisation to higher polygons}
Summing up the divergent contributions of \eqref{eqn:result-insertion-interaction-term} \eqref{eqn:result-insertion-kinetic-term-vertex}, and inserting
them into the dilatation Ward identity \eqref{eqn:dilatation-Wardidentitiy},
we obtain
\begin{equation}\label{eqn:dilatation-wi}
\mathbb{D} \langle W_4 \rangle^{(2)} = 
-\left(\frac{N}{k}\right)^2  \ln(2) \left( \sum_{i=1}^{4} 1 \right)  + \mathcal{O}(\epsilon)\,,
\end{equation}
where we have written the factor $4$ as $(\sum_{i=1}^{4} 1)$ to emphasise its origin
from the sum of four vertex-type diagrams.
Note that only the divergent part of the vertex-diagram was required here.
Summing up \eqref{eqn:special-conformal-wi-vertex-insertion}  and 
\eqref{eqn:special-conformal-wi-vertex-insertion-in-propagator}, and inserting them into the 
special conformal Ward identity \eqref{eqn:special-conformal-Wardidentitiy}, we obtain
\begin{equation}\label{eqn:special-conf-wi}
\mathbb{K}^\nu \langle W_4 \rangle^{(2)}
= -2 \left( \frac{N}{k} \right)^2  \ln(2) \, \sum_{i=1}^{4} x_i^\nu + \mathcal{O}(\epsilon)\,.
\end{equation}
Let us now explain how these equations can be generalised from $n=4$ to arbitrary $n$.
In our two-loop computation, we found that the only diagrams contributing 
to \eqref{eqn:dilatation-wi} and \eqref{eqn:special-conf-wi}
are those producing poles in $\epsilon$.
The mechanism for how these poles are generated was described in section \ref{sect:vertex-diagram}, see in
particular Figure \ref{fig:divergence-vertex}. It is clear that for $n>4$ cusps, the same type of vertex diagram
will produce the divergent terms. Although those diagrams will depend on one further
kinematical variable w.r.t. the four-point case, this dependence cannot change the (leading) 
UV pole $\epsilon^{-1}$ of the diagrams. Since there are $n$ diagrams of this type at
$n$ points, we expect
\begin{equation}\label{eqn:dilatation-wi-n}
\mathbb{D} \langle W_n \rangle^{(2)} = 
-\left(\frac{N}{k}\right)^2  \ln(2) \left( \sum_{i=1}^{n} 1 \right)  + \mathcal{O}(\epsilon)\,,
\end{equation}
and
\begin{equation}\label{eqn:special-conf-wi-n}
\mathbb{K}^\nu \langle W_n \rangle^{(2)}
= -2 \left( \frac{N}{k} \right)^2  \ln(2) \, \sum_{i=1}^{n} x_i^\nu + \mathcal{O}(\epsilon)\\,.
\end{equation}
We will now proceed to discuss the solution of these Ward identities and
compare them to the result of the two-loop computation of the tetragon Wilson loop in section \ref{sect:CS-2loop-results}.

\subsection{Solution to the anomalous conformal Ward identities}
Using
$  \mathbb{D}\left( x_{ij}^2\right) = 2 x_{ij}^2 $
it is clear that the most general solution to the dilatation Ward identity \eqref{eqn:dilatation-wi-n} is 
\begin{align}\label{eqn:solution-dilatation-ward-identity-n}
\langle W_n \rangle^{(2)} = -\left(\frac{N}{k}\right)^2  \left[  \frac{\ln(2)}{4}\sum_{i=1}^4\frac{(-x_{i,i+2}^2 \tilde \mu^2 )^{2\epsilon}}{\epsilon} + f_{n}\left(\frac{x_{ij}^{2}}{x_{kl}^2}\right)\right] + O(\epsilon) \,,
\end{align}
where $f$ is an arbitrary function of dimensionless variables and we recall that  
$\tilde{\mu}^2 = \mu^2 \pi e^{\gamma_{E}}$. 
Of course, this is exactly what we expect from \eqref{eqn:wilson-loop-regularisation}. 

The result for the special conformal Ward identity is more interesting. 
Plugging \eqref{eqn:solution-dilatation-ward-identity-n} into the special
conformal Ward identity \eqref{eqn:special-conf-wi}
and using  $\mathbb{K}^\nu \ln (x_{kl}^2) = 2(x_k + x_l)^\nu $,
it is easy to see that the function $f_{n}$ is allowed to
depend on conformally invariant cross-ratios only, i.e. $f_{n}(x_{ij}^2 / x_{kl}^2 )$ = $g(u_{abcd})$.
Therefore we finally have
\begin{align}\label{eqn:solution-special-conformal-ward-identity-xij}
\langle W_n \rangle^{(2)} &=  -\left(\frac{N}{k}\right)^2  \left[ \frac{\ln(2)}{4}\sum_{i=1}^n\frac{(-x_{i,i+2}^2 \tilde \mu^2 )^{2\epsilon}}{\epsilon} + g_{n}(u_{abcd}) + \mathcal{O}(\epsilon) \right]\,.
\end{align}
In the four-point case, there are no non-vanishing cross-ratios,
and therefore in particular $g_{4}$ must be a constant.
This is in agreement with \eqref{eqn:result-wilson-loop-two-loop} and thus 
represents an independent check of the direct perturbative computation,
including its finite part (recall that deriving the Ward identity does not rely 
on the finite parts of the direct perturbative computation). 
So, even though the result for the vertex diagram \eqref{eqn:result-vertex} was 
obtained numerically, its functional form is an analytical result, since we know the 
analytical expression for the ladder diagram and the sum of vertex and ladder diagram 
through the solution to the anomalous conformal Ward identity.

%%%%%%%%%%%%%%%%%%%%%%%%%%%%%%%%%%%%%%%%%%%%%%%%%%%%%%%%%%%%%%%%%%%%%%%%%%%%%%%%
\section{Two loops: ABJM theory}
Here we explain how the results are modified in ABJM theory. We use the Wilson loop operator proposed in \cite{Drukker:2008zx}
\begin{align}\label{eqn:Wilson-loop-ABJM}
\langle W(A,\hat A) \rangle =  \frac{1}{2N} \left\langle   \tr \mathcal{P}\exp \left(i \oint_\mathcal{C} A_\mu dz^\mu \right) + \hat\tr \mathcal{P}\exp \left(i \oint_\mathcal{C} \hat{A}_\mu dz^\mu \right)  \right\rangle \,.
\end{align}
Note that the sign(s) in the exponent(s) in \eqref{eqn:Wilson-loop-ABJM} are correlated to corresponding signs in the Lagrangian by the requirement of
gauge invariance, see Appendix \ref{app:Conventions}.

\subsection{Gauge field contributions}
In ABJM theory there is a second copy of the gauge field $\hat A_\mu$ with opposite sign in the Lagrangian \eqref{eqn:ABJM-Lagrangian}. 
Up to a sign, the gauge field contributions for both gauge groups are identical at one loop,
\begin{align}
\langle W \rangle^{(1)}_A = - \langle W \rangle^{(1)}_{\hat A}\,,
\end{align}
due to the different sign of the propagator for the second gauge field, $\langle A_\mu A_\nu \rangle = - \langle \hat A_\mu \hat A_\nu \rangle $. 
Thus, at one loop the diagrams cancel. This does not differ from the result of pure Chern-Simons theory, since the expectation value at one loop vanishes, 
as we found in section \ref{sec:one-loop} for $n=4,6$.

At two loops, however, the sign has no effect, since the two-gluon diagram contains an even number of propagators 
and in the vertex diagram we have to take into account the sign of the interaction term as well. 
Therefore, the two-loop diagrams are identical
\begin{equation}
\langle W \rangle^{(2)}_A = \langle W \rangle^{(2)}_{\hat A}
\end{equation}
Thus, up to two loops, the expectation value for pure gauge field contributions is the same in ABJM theory and Chern-Simons theory
\begin{align}
\langle W(A,\hat A) \rangle_{\text{gauge fields}}= \langle W(A) \rangle_{\text{CS}}\,.
\end{align}

\subsection{Matter contributions}\label{sec:matter-contributions}
In pure Chern-Simons theory the one-loop correction to the gauge field propagator is zero, 
since the contributions of gauge fields and ghosts exactly cancel against each other, see \eqref{eqn:gluon-ghost-loops}. 

In ABJM theory we have to take into account fermionic and bosonic matter in the loop. 
This gauge field self energy has been calculated in 
\cite{Gaiotto:2007qi,Bak:2008cp,Drukker:2008zx} and the corrected 
propagator reads\footnote{Recall that we absorbed the regularisation scale into the coupling constant: $k \rightarrow \mu^{-2 \epsilon} k $}

\begin{align}\label{eqn:one-loop-correctd-gluon-prop-tot-der}
G_{\mu\nu}^{(1)}(x) 
&= \left(\frac{2\pi}{k}\right)^2  \frac{N \delta^I_I}{8} \frac{\Gamma(1-\frac{d}{2}) \Gamma(\frac{d}{2})^2}{\Gamma(d-1)\,\pi^d}    \left(  \frac{\Gamma(d-2)}{\Gamma(2-\frac{d}{2})} \frac{\eta_{\mu\nu}}{(-x^2)^{d-2}} -\partial_\mu \partial_\nu\left( \frac{\Gamma(d-3)}{\Gamma(3-\frac{d}{2})} \frac{1}{4}  \frac{1}{(-x^2)^{d-3}}  \right) \right)\,,
\end{align}
for details see Appendix \ref{sec:one-loop-correction-propagator}.
%Evaluating the derivatives, we get
%\begin{align}
% G_{\mu\nu}^{(1)}(x) 
% &= \left(\frac{2\pi}{k}\right)^2  \frac{N \delta^I_I}{8} \frac{\Gamma(1-\frac{d}{2}) \Gamma(\frac{d}{2})^2}{\Gamma(d-1)\,\pi^d}    \left( \frac{\Gamma(d-1)}{\Gamma(3-\frac{d}{2})}  \frac{x_\mu x_\nu}{(x^2)^{d-1}}   + \underbrace{\frac{\Gamma(d-2)}{\Gamma(3-\frac{d}{2})} \frac{\eta_{\mu\nu}\left(-\frac{1}{2} + (2-\frac{d}{2}) \right)}{(x^2)^{d-2}}}_{\mathcal{O}(\epsilon)} \right)\,.
%\end{align}
%One can either use this form of the propagator or take \eqref{eqn:one-loop-correctd-gluon-prop-tot-der} and 
%drop the derivative term to calculate the expectation value.  Both calculations yield the same divergent and finite terms.
%
%We illustrate the calculation for the latter case, i.e. we drop the derivative term and use the propagator
We can drop the derivative term in \eqref{eqn:one-loop-correctd-gluon-prop-tot-der}  (it
would not contribute to the gauge-invariant Wilson loop) and instead use the propagator
\begin{align}
G_{\mu\nu}^{(1)}(x) 
&= -\frac{1}{N}\left(\frac{N}{k} \right)^2 \pi^{2-d} \Gamma \left( \frac{d}{2}-1 \right)^2   \frac{\eta_{\mu\nu}}{(-x^2)^{d-2}} 
%&= -\frac{1}{N}\left(\frac{N}{k} \right)^2 \left( (4 \pi e^{\gamma_E} )^{2 \epsilon} + \frac{\pi^2}{2}\epsilon^2 + \mathcal{O}(\epsilon^3) \right) \frac{\eta_{\mu\nu}}{(x^2)^{d-2}}
\,,
\end{align}
which up to two small differences is  the tree level $\mathcal{N}=4$ SYM gluon propagator. 
The first difference is a trivial prefactor, and the second is that since we are at two loops, 
the power of $1/x^2$ is $1-2 \epsilon$ here,
as opposed to $1-\epsilon$ in the one-loop computation in $\mathcal{N}=4$ SYM.
Thus it is clear that the results will be very similar to the expectation value of the Wilson loop in $\mathcal{N}=4$ SYM.
The corresponding calculation in $\mathcal{N}=4$ SYM was carried out in \cite{Drummond:2007aua}
and we briefly summarise the results.

\begin{figure}[t]
\centering
\subfloat[]{\includegraphics[width=.15 \textwidth]{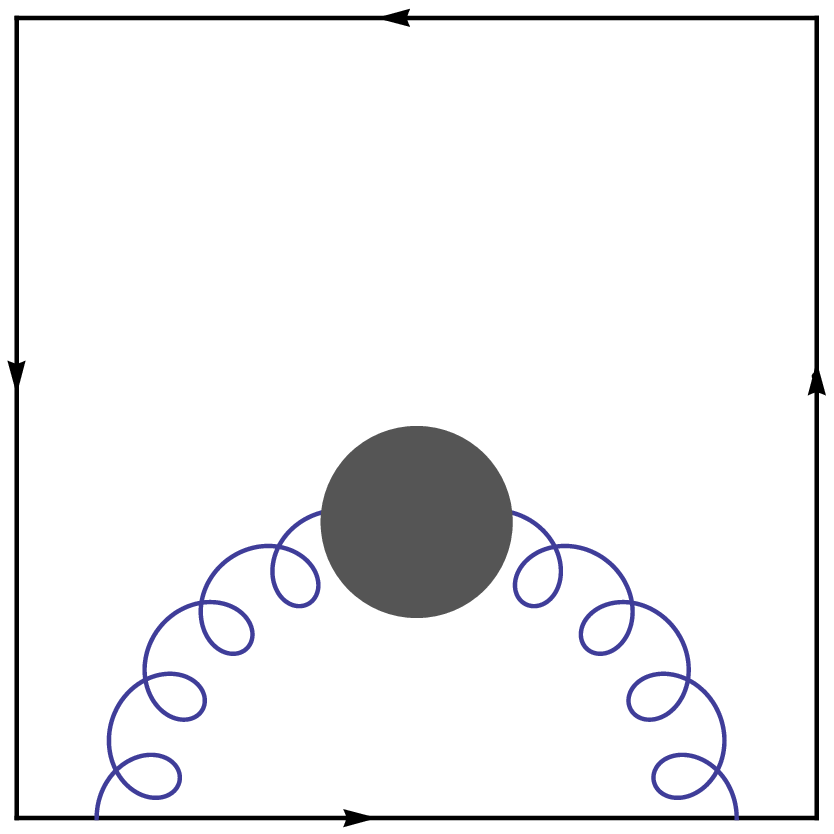}\label{fig:general-diagram-propagator-corrections-same-edge}}~~~~~
\subfloat[]{\includegraphics[width=.15 \textwidth]{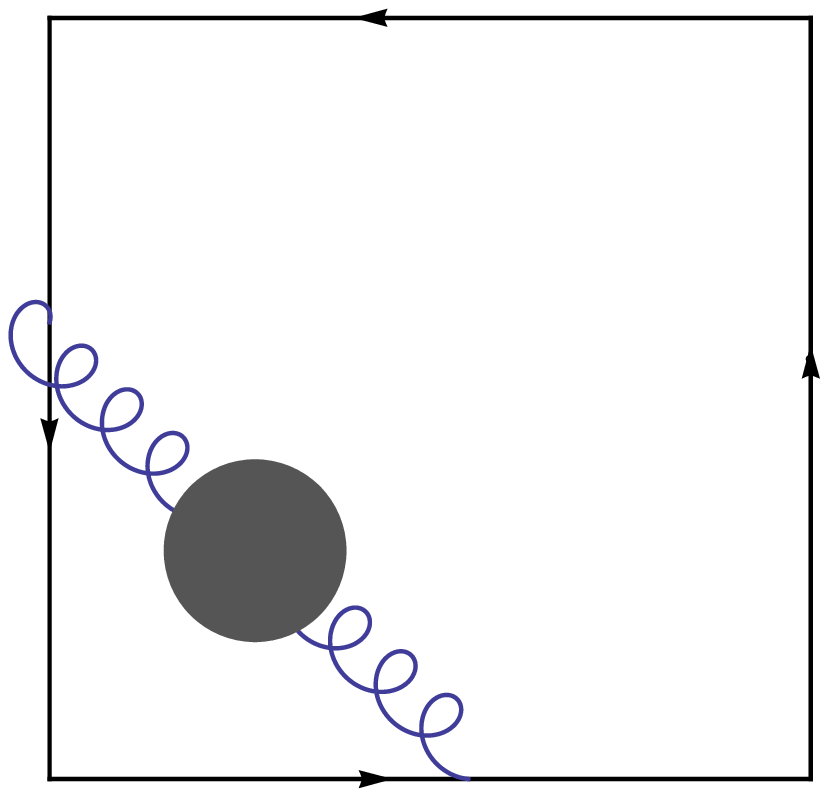}\label{fig:general-diagram-propagator-corrections-adjacent}}~~~~~
\subfloat[]{\includegraphics[width=.15 \textwidth]{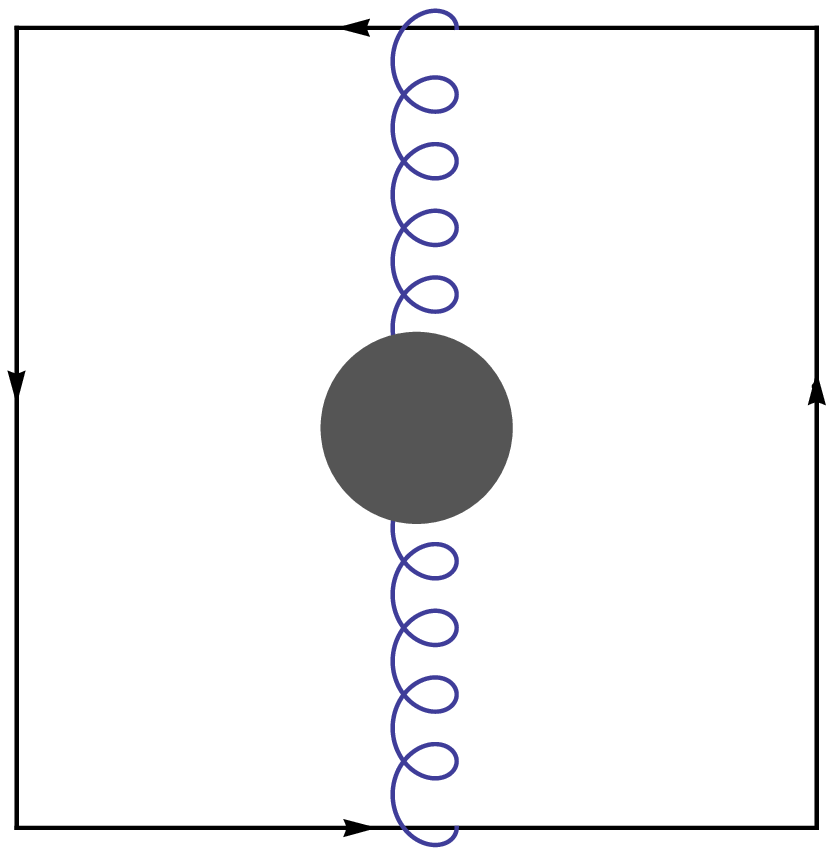}\label{fig:general-diagram-propagator-corrections-non-adjacent}}~~~~~~~~~~
\caption{Examples for the three classes of diagrams involving the gauge field self energy: diagrams with (a) a propagator connecting the same edge (vanishing), (b) propagator stretching between adjacent edges (divergent), (c) propagator stretching between non-adjacent edges (finite). These diagrams have the same structure as the 1-loop diagrams in $\mathcal{N}=4$ SYM.
}
\label{fig:self-energy-diagrams}
\end{figure}
As in $\mathcal{N}=4$ SYM we have three classes of diagrams  shown in figure \ref{fig:self-energy-diagrams}. 
Diagram \ref{fig:general-diagram-propagator-corrections-same-edge} vanishes due to the light-likeness of the edges, 
whereas \ref{fig:general-diagram-propagator-corrections-adjacent} yields a divergent, and 
\ref{fig:general-diagram-propagator-corrections-non-adjacent} yields a finite contribution.
We have
\begin{align}\label{eqn:general-expression}
\langle W_4 \rangle^{(2)}_{\text{matter}} &= \frac{i^2}{N}\tr \int_{z_i>z_j} dz_i^\mu dz_j^\nu \langle A_\mu A_\nu \rangle^{(1)}\\ \nonumber
&=- N \sum_{i>j} \int ds_i ds_j p_i^\mu  p_j^\nu G_{\mu\nu}^{(1)}(z_i-z_j) \\ \nonumber
&= \left( \frac{N}{k}\right)^2 \left( (4 \pi e^{\gamma_E} )^{2 \epsilon} + \frac{\pi^2}{2}\epsilon^2 + \mathcal{O}(\epsilon^3) \right) \sum_{i>j} I_{ij}
\end{align}
where $I_{ij}= \int ds_i ds_j\, p_i \cdot p_j (-(z_i-z_j)^2)^{2-d}$.

There are four diagrams  $I_{i+1,i}$ of the type shown in fig. \ref{fig:general-diagram-propagator-corrections-adjacent}. 
It is sufficient to compute one of them, as the others can be obtained by the replacement $i \rightarrow i+1$.  Setting e.g. $i=2,j=1$, we have
\begin{align}
I_{21} &=  
%\int_0^1 ds_2 ds_1 \frac{p_2 \cdot p_1}{\left((z_2-z_1)^2\right)^{d-2}} = 
-\frac{1}{2} (-x_{13}^2)^{3-d} \int_0^1 ds_2 ds_1 \frac{1}{\left(\bar{s}_1  s_2\right)^{1-2\epsilon}}
= -\frac{1 }{8} \frac{(-x_{13}^2)^{2\epsilon}}{\epsilon^2} 
\end{align}
Furthermore, there are two finite diagrams $I_{i+2,i}$ of the type shown in fig. \ref{fig:general-diagram-propagator-corrections-non-adjacent}. 
Setting $d=3$ and taking $i=3, j=1$ we have
\begin{equation}
I_{31} 
%=   \int_0^1 ds_3 ds_1 \frac{p_3 \cdot p_1}{(z_3-z_1)^2} 
= \frac{1}{2} {\int_0^1 ds_3 ds_1 \frac{x_{13}^2+x_{24}^2}{ x_{13}^2 \bar{s}_1 \bar{s}_3 + x_{24}^2 s_1 s_3  }} 
=  \frac{1}{4}\left[ \ln^2\left(\frac{x_{13}^2}{x_{24}^2} \right)+ \pi^2 \right]\,.
\end{equation}
Taking the sum over all contributions we obtain
\begin{equation}
\sum_{i>j} I_{ij} = \frac{1}{4}\left[  -\frac{(-x_{13}^2)^{2\epsilon}}{\epsilon^2} -\frac{(-x_{24}^2)^{2\epsilon}}{\epsilon^2}  + 2  \ln^2\left(\frac{x_{13}^2}{x_{24}^2} \right) + 2 \pi^2 \right]\,,
\end{equation}
and thus the full matter part is
\begin{equation}\label{eqn:result-matter-part}
\langle W_4 \rangle^{(2)}_{\text{matter}}
= -\frac{1}{4}\left( \frac{N}{k}\right)^2 \left[ \frac{(-x_{13}^2 4 \pi e^{\gamma_E}\mu^2 \, )^{2\epsilon}}{\epsilon^2} +
\frac{(- x_{24}^2 4 \pi e^{\gamma_E}\mu^2 \,)^{2\epsilon}}{\epsilon^2}  - 2  \ln^2\left(\frac{x_{13}^2}{x_{24}^2}\right) - \pi^2 + \mathcal{O}(\epsilon) \right]\,,
\end{equation}
where we have restored the regularisation scale $\mu^2$.
The full result is obtained by adding the CS part  \eqref{eqn:result-wilson-loop-two-loop}
%\begin{equation}
%\langle W_4 \rangle^{(2)}_{\text{CS}} =  - \frac{1}{4}\left(\frac{N}{k}\right)^2  \left( 2\ln(2)\frac{(\pi %e^{\gamma_E} \mu^2 s )^{2\epsilon}+
%(\pi e^{\gamma_E} \mu^2  t )^{2\epsilon}}{\epsilon} + (a_6 - 8 \ln(2) - \pi^2) \right)\,.
%\end{equation}
to \eqref{eqn:result-matter-part}
and the result can be rewritten in a form in which the $\epsilon^{-1}$ terms cancel
\begin{align}
\langle W_4 \rangle^{(2)}_{\text{ABJM}} = -\frac{1}{4}\left( \frac{N}{k}\right)^2 \left[   \frac{(-{\mu^\prime}^2 \, x_{13}^2)^{2\epsilon}}{\epsilon^2} +
\frac{(-{\mu^\prime}^2 \, x_{24}^2)^{2\epsilon}}{\epsilon^2}  - 2  \ln^2\left(\frac{x_{13}^2}{x_{24}^2}\right)  -const. + \mathcal{O}(\epsilon) \right] \,,
\end{align}
where ${\mu^\prime}^2= 8 \pi e^{\gamma_E}\mu^2 $ and  $const.= 8 \ln (2) + 20 \ln^2(2)+2\pi^2-a_6$ and where $a_6$ is given in \eqref{eqn:a6} and fits the value $a_6 = -\frac{2}{3}\pi^2 + 16 \ln(2) + 8 \ln^2 (2)$ such that $const.= \frac{8}{3}\pi^2 + 12 \ln^2(2)-8 \ln(2)$.\footnote{Note that the $\log(2)$ term has a different transcendentality than the other terms. It stems from the vertex diagrams of the pure CS part \eqref{eqn:result-wilson-loop-two-loop} of the Wilson loop.}
This is the result quoted in the introduction
\eqn{here}.

%%%%%%%%%%%%%%%%%%%%%%%%%%%%%%%%%%%%%%%%%%%%%%%%%%%%%%%%%%%%%%%%%%%%%%%%%%%%%%%%
\subsection*{Acknowledgements}

We are grateful to N.~Beisert, D.~Berman, 
H.~Dorn, N.~Drukker, C.~Grosse Wiesmann, T.~Mc\-Lough\-lin,
P.~Nair, D.~Sorokin and T.~Schuster for valuable
comments and discussions. J.H. is grateful to the Institute for Advanced Study, Princeton,
for hospitality during the final stage of this work.
This work was supported by the Volkswagen Foundation.

%%%%%%%%%%%%%%%%%%%%%%%%%%%%%%%%%%%%%%%%%%%%%%%%%%%%%%%%%%%%%%%%%%%%%%%%%%%%%%%%
%%%%%%%%%%%%%%%%%%%%%%%%%%%%%%%%%%%%%%%%%%%%%%%%%%%%%%%%%%%%%%%%%%%%%%%%%%%%%%%%
%%%%%%%%%%%%%%%%%%%%%%%%%%%%%%%%%%%%%%%%%%%%%%%%%%%%%%%%%%%%%%%%%%%%%%%%%%%%%%%%
\appendix

%%%%%%%%%%%%%%%%%%%%%%%%%%%%%%%%%%%%%%%%%%%%%%%%%%%%%%%%%%%%%%%%%%%%%%%%%%%%%%%%
%%%%%%%%%%%%%%%%%%%%%%%%%%%%%%%%%%%%%%%%%%%%%%%%%%%%%%%%%%%%%%%%%%%%%%%%%%%%%%%%
\section{Conventions}
\label{app:Conventions}
An $n$-sided polygon can be defined by $n$ points $x_i$ ($i=1,..., n$), with the edge $i$ being the line connecting $x_i$ and $x_{i+1}$. Defining
\begin{equation}
p_i^{\mu}=x^{\mu}_{i+1}-x^{\mu}_i
\end{equation}
and parametrising the position $z^{\mu}_i$ on edge $i$ with the parameter $s_i \in [0,1]$ we have
\begin{equation} \label{eqn:z-parametrization}
z^{\mu}_i(s_i)= x^{\mu}_i + p^{\mu}_i s_i \,.
\end{equation}
Furthermore, we use the notation
\begin{equation}
\epsilon(p,q,r) = \epsilon_{\mu\nu\rho} p^{\mu} q^{\nu} r^{\rho} \qquad 
 \text{and} \qquad \bar{s}_i= 1- s_i.
\end{equation}
One can easily check that with the definition $x^{\mu}_{ij} = x^{\mu}_i-x^{\mu}_j $
\begin{equation}\label{eqn:scalprodx_ijx_mn}
2\, x_{ij}\cdot x_{mn}= x^2_{in}+x^2_{jm} - x^2_{im} - x^2_{jn}
\end{equation}
We consider 3-dimensional Minkowski space with metric
\begin{equation}
\eta_{\mu\nu}= \text{diag } (1,-1,-1) \,.
\end{equation}
Using \eqref{eqn:scalprodx_ijx_mn} we can rewrite the scalar products
\begin{equation} \label{eqn:products:p_ip_m}
2 p_i \cdot p_j  = x^2_{i,j+1} +  x^2_{i+1,j} -  x^2_{i,j} -  x^2_{i+1,j+1} \,.
\end{equation}
Furthermore, using the definition \eqref{eqn:z-parametrization}, one can easily show that
\begin{equation}\label{eqn:z_i-z_j^2}
(z_i-z_j)^2=x^2_{ij}\bar{s}_i\bar{s}_j + x^2_{i+1,j} s_i \bar{s}_j + x^2_{i,j+1}\bar{s}_i s_j+x^2_{i+1,j+1}s_i s_j\, .
\end{equation}

\subsection{Cherns-Simons theory and the Wilson loop operator}
The Chern-Simons action
\begin{equation}\label{eqn:pure-CS-Lagrangian}
S_{\text{CS}}=  \frac{k}{4 \pi} \int d^dx\, \epsilon^{\mu\nu\rho}\, \tr \left( A_\mu \partial_\nu A_\rho - \frac{2}{3} i A_\mu A_\nu A_\rho \right) 
\end{equation}
and the Wilson loop operator
\begin{align}
W[\,\mathcal{C}\,] = \frac{1}{N}\tr \mathcal{P} \exp \left( i \oint_{\mathcal{C}} A_\mu dz^\mu \right)\,.
\end{align}
are invariant\footnote{More precisely, the action is invariant under infinitesimal transformations $g(x)=1 + i \alpha(x)$ and transforms like $ S \rightarrow S^\prime + (2\pi k ) \delta S$, where $\delta S=- \frac{1}{24\pi^2}\int d^dx\, \epsilon^{\mu\nu\rho}\tr \left( (\partial_\mu g^{-1})g(\partial_\nu g^{-1})g(\partial_\rho g^{-1})g \right)$ under finite transformations. Since $\delta S$ takes integer values, $\exp ({i S})$ is invariant under large gauge transformations for $k \in \mathbb{N}$.} under $SU(N)$ gauge transformations
\begin{align}
A_\mu \rightarrow A_\mu^\prime= g(x)\left(A_\mu + i \partial_\mu \right)g^{-1}(x)\,, \qquad g(x) \in SU(N)
\end{align}
if the path ordering\footnote{There are different conventions on the path ordering. They are equivalent, it is however important to choose the sign in the Wilson loop such that it is gauge invariant. Reversing the sign in the exponent of the Wilson loop does not just reverse the integration contour, due to the path ordering.} is defined as
\begin{align}\label{eqn:definition-path-ordering}
& \mathcal{P} \left(A_\mu\left( z(s) \right) A_\nu\left( z(s^\prime) \right) \right) = A_\mu\left( z(s) \right) A_\nu\left( z(s^\prime) \right) \qquad \text{for}\quad  s>s^\prime\,, \\ \nonumber
& \mathcal{P} \left(A_\mu\left( z(s) \right) A_\nu\left( z(s^\prime) \right) \right) = A_\nu\left( z(s^\prime) \right) A_\mu\left( z(s) \right) \qquad \text{for}\quad  s<s^\prime 
\end{align}
and $s \in (0,1)$ parametrises the path along the curve $\mathcal{C}$. The path ordered exponential in the Wilson loop operator then has the expansion
\begin{align}\label{eqn:Wilson-Loop-after-path-ordering}
\mathcal{P} \exp{\left(i \oint_{\mathcal{C}} A_{\mu} dz^{\mu}\right)}&=  1_{\text{N} \times \text{N}} + i \oint_{\mathcal{C}}dz^{\mu}_i A_{\mu}  + (i)^2 \oint_{\mathcal{C}}dz^{\mu}_i \int^{z_i} dz_j^{\nu} A_{\mu} A_{\nu} \\ \nonumber
&\phantom{=}+ (i)^3 \oint_{\mathcal{C}}dz^{\mu}_i \int^{z_i} dz_j^{\nu}\int^{z_j} dz_k^{\rho} A_{\mu} A_{\nu} A_{\rho} \\ \nonumber
&\phantom{=}+ (i)^4\oint_{\mathcal{C}}dz^{\mu}_1 \int^{z_i} dz_j^{\nu}\int^{z_j} dz_k^{\rho}\int^{z_k} dz_l^{\sigma} A_{\mu} A_{\nu} A_{\rho}A_{\sigma} +~...\,.
\end{align}

Quantising the theory with the standard Fadeev-Popov procedure yields the gauge fixing and ghost action
\begin{equation}
\mathcal{S}_{\text{g.f.}}=  \frac{k}{4\pi}\int d^dx\,   \tr \left(\frac{1}{\xi} \left( \partial^\mu A_\mu \right)^2 +  \bar{c} \left( \partial^\mu D_\mu \right)c \right)
\end{equation}
where $ D_\mu c= \partial_\mu c +i [A_\mu, c\, ]$. In Landau gauge $(\xi=0)$ the gauge field
propagator reads
\begin{align}%\label{eqn:gluon-prop}
\langle \left(A_\mu\right)_{ij}(x) \left(A_\nu\right)_{kl}(y) \rangle &= \delta_{il} \delta_{jk} \frac{1}{k}\left( \frac{\Gamma\left(\frac{d}{2}\right)}{\pi^{\frac{d-2}{2}}}\right) \epsilon_{\mu\nu\rho}\frac{(x-y)^\rho}{\left(-(x-y)^2\right)^{\frac{d}{2}}}
\end{align}
where we have rescaled the coupling constant $k \rightarrow \mu^{-2\epsilon}k$ and restore the dependence on the regularisation scale $\mu$ only in the final results.
The ghost propagator is
\begin{equation}
\langle c(x) \bar{c}(y) \rangle =  \delta_{il} \delta_{jk} \frac{1}{k}\left( \frac{\Gamma\left(\frac{d}{2}-1\right)}{\pi^{\frac{d-2}{2}}}\right)\frac{1}{\left(-(x-y)^2\right)^{\frac{d}{2}-1}}\,.
\end{equation}
Note that the gauge field propagator is related to the ghost propagator by
\begin{equation}\label{eqn:ghost-gluon-propagator-relation}
\langle \left(A_\mu\right)_{ij}(x) \left(A_\nu\right)_{kl}(y) \rangle = \frac{1}{2}\epsilon_{\mu\nu\rho} \partial^\rho \langle c(x) \bar{c}(y) \rangle \,,
\end{equation}
which can be used to see the cancellation of gauge field 
and ghost loop contributions to the one-loop gauge field propagator in a simple way.

The different conventions found in the literature on Chern-Simons theory deserve a short comment. One can show, that 
\begin{align}
\mathcal{S}&= \frac{k}{4 \pi} \int d^dx\, \tr \left( A_\mu \partial_\nu A_\rho - \frac{2}{3} s A_\mu A_\nu A_\rho \right) \epsilon^{\mu\nu\rho}\,,\\
W(\mathcal{C}) &= \frac{1}{N}\tr \mathcal{P} \exp \left( s \oint_{\mathcal{C}} A_\mu dz^\mu \right)
\end{align}
are invariant  ({In the sense mentioned above}) under gauge transformations 
\begin{equation}
A_\mu \rightarrow A_\mu^\prime = g(x)\left(A_\mu - \frac{1}{s} \partial_\mu \right)g^{-1}(x)
\end{equation}
where s is a real or imaginary parameter. I.e. the sign in the Wilson loop and the Lagrangian are correlated through gauge invariance. Taking a hermitian gauge field $(A_\mu)^\dagger=A_\mu$ we can choose $s=i$, which corresponds to the choice used throughout this document.\\

All other conventions found in the literature can be obtained by rescaling the gauge field $A_\mu \rightarrow -A_\mu$, $A_\mu \rightarrow i A_\mu$ etc. Note however, that this changes factors in the Lagrangian, the gauge transformation, the covariant derivative and the Wilson loop. A sign difference in the Wilson loop only may also be due to a different definition of the path ordering \eqref{eqn:definition-path-ordering}.

\subsection{Lagrangian of ABJM theory}
The action of ABJM theory is
\begin{equation}\label{eqn:ABJM-Lagrangian}
\mathcal{S}_{\text{ABJM}}=  \mathcal{S}_{\text{CS}}+\mathcal{S}_{\text{g.f.}}+\mathcal{\hat{S}}_{\text{CS}}+\mathcal{\hat{S}}_{\text{g.f.}}+\mathcal{S}_{\text{matter}}
\end{equation}
where $\mathcal{\tilde{S}}$ is obtained from $\mathcal{S}$ by replacing $A_\mu$ with the gauge field in the anti-fundamental representation $\hat{A}_\mu$ and letting $k\rightarrow -k$. Explicitly, we have
\begin{align}
S_{\text{CS}} + \hat{S}_{\text{CS}} & = \frac{k}{4\pi}\,\int d^3x\, \varepsilon^{\mu\nu\rho} \,\Bigl [\,
\Tr (A_\mu\partial_\nu A_\rho-\frac{2}{3}i\,A_\mu A_\nu A_\rho)- 
\Tr (\hat A_\mu\partial_\nu \hat A_\rho-\frac{2}{3}i\, \hat A_\mu \hat A_\nu \hat A_\rho)\, \Bigr ]  \\
S_\text{gf}+\hat S_\text{gf} & = \frac{k}{4\pi}\, \int d^3 x\, \Bigl [\,\frac{1}{\xi}\, \Tr(\partial_\mu A^\mu)^2
-\Tr(\partial_\mu \bar c\, D_\mu c) - \frac{1}{\xi}\, \Tr(\partial_\mu \hat A^\mu)^2
+\Tr(\partial_\mu \bar{ \hat c}\, D_\mu \hat c) \, \Bigr ] \\ \label{eqn:ABJM-part-Lagrangian}
S_\text{matter} & = \int d^3 x\, \Bigl [\, \Tr(D_\mu\, C_I\, D^\mu {\bar C}^{I}) + i\, \Tr(\bar\psi^I
\, \slsh{D}\, \psi_I)\, \Bigr ] + S_\text{int}
\end{align}
The field content consists of two $U(N)$ gauge fields $(A_\mu)_{ij}$ and
$(\hat A_\mu)_{\hi\hj}$, the complex fields $(C_I)_{i\hi}$ and $({\bar C}^{I})_{\hi i}$ as
well as the fermions $(\psi_I)_{\hi i}$ and $({\bar\psi}^{I})_{i \hi }$ in 
the $({\bf N},{\bf \bar N})$ and $({\bf \bar N},{\bf N})$ of $U(N)$ respectively,
$I=1,2,3,4$ is the $SU(4)_R$ index. We employ the covariant gauge fixing function 
$\partial_\mu A^\mu$ for both gauge fields and have two sets of ghosts $(\bar c,c)$ and
$(\bar{\hat c},\hat c)$. $S_\text{{int}}$ are the sextic scalar potential and $\Psi^2C^2$ Yukawa type potentials spelled out in \cite{Aharony:2008ug}. The covariant derivative is given by $D_\mu c = \partial_\mu c + i [A_\mu, .]$. It's action on $\Psi^I, C^I$ is given in \cite{Drukker:2008zx} and not needed here.

%%%%%%%%%%%%%%%%%%%%%%%%%%%%%%%%%%%%%%%%%%%%%%%%%%%%%%%%%%%%%%%%%%%%%%%%%%%%%%%%
\section{Details of the two-loop calculation}
\label{app:Details}

\subsection{Vertex diagram}\label{app:Vertex}
Using $\epsilon(p_1,p_2,z_{12})=\epsilon(p_3,p_2,z_{32})=0$ the first line in \eqref{eqn:vertex-Iijk} can be rewritten as
\begin{equation}
I_{321}=\int d^3s_{1,2,3} \epsilon(p_1,p_2,\partial_{z_1})\epsilon(p_3,p_2,\partial_{z_3}) \int d^d w \frac{(d-2)^{-2}}{|w|^d|w-z_{12}|^{d-2}|w-z_{32}|^{d-2}}\,.
\end{equation}
We begin by introducing Feynman parameters,
\begin{equation}
\int d^d w \frac{1}{|w|^d|w-z_{12}|^{d-2}|w-z_{32}|^{d-2}} = \int [d\beta]_3 \int d^d w \frac{1}{\left(-(w-\beta_1 z_{12}-\beta_3 z_{32})^2 + \Delta\right)^{(3d-4)/2}}\,,
\end{equation}
where
\begin{equation}
\int [d\beta]_3 = \int_0^1 d\beta_1d\beta_2d\beta_3 (\beta_1\beta_2\beta_3)^{(d-2)/2-1}\beta_2 \delta(\sum_i \beta_i-1) \frac{\Gamma(\frac{3d}{2}-2)}{\Gamma(\frac{d}{2})\Gamma(\frac{d}{2}-1)^2}
\end{equation}
and
\begin{equation}\label{eqn:Delta-denominator-vertex}
\Delta= 2 \beta_1\beta_3 (z_{12}\cdot z_{32})- z_{12}^2 \beta_1 \bar{\beta}_1- z_{32}^2 \beta_3 \bar{\beta}_3\,.
\end{equation}
Shifting the integration contour $w\rightarrow l=w - \beta_1 z_{12}-\beta_3 z_{32}$ we have a standard integral
\begin{align}
\int d^d l  \frac{1}{[l^2 - \Delta]^{n}}=(-1)^n  i \pi^{d/2}\frac{\Gamma(n-\frac{d}{2})}{\Gamma(n)}\left(\frac{1}{\Delta}\right)^{n-\frac{d}{2}}\,.
\end{align}
Thus we get
\begin{equation}
I_{321}= \frac{c_1 }{(d-2)^2} \int [d\beta]_3d^3s_{1,2,3} \epsilon(p_1,p_2,\partial_{z_1})\epsilon(p_3,p_2,\partial_{z_3})\frac{1}{\Delta^{d-2}} \,,
\end{equation}
where
%\begin{align}
$c_1= i \pi^{\frac{d}{2}}{\Gamma\left(d-2\right)}/{\Gamma\left({3d}/{2}-2\right)} $.
%\end{align}
Evaluating the action of the derivatives  and abbreviating $x_{13}^2=s, x_{24}^2=t$ we obtain
\begin{align}\label{eqn:I123}
I_{321}&= c_2 s t  \int_0^1 d^3s_{1,2,3} d^3\beta_{1,2,3}(\beta_1\beta_2\beta_3)^{(d-2)/2} \delta(\sum_i \beta_i-1) \left( \frac{ 1}{\Delta^{d-1}}- 2\frac{(d-1)}{\Delta^{d}}\beta_1\beta_3 \bar{s}_1 s_3 (s+t) ) \right)
\end{align}
where 
%\begin{equation}\label{eqn:c2}
$c_2={i \pi^{{d}/{2}}}{\Gamma(d-1)}/( 8 {\Gamma^{3}({d}/{2})})
%=i + \mathcal{O}(\epsilon)
$
%\end{equation}
and both terms are separately symmetric under $s \leftrightarrow t$. 
Performing the change of variables 
\begin{equation}
\beta_1= x y\,,\qquad \beta_2= \bar{x} y\,,\qquad\beta_3= \bar{y}\,,\qquad \sum_i \beta_i=1\,\qquad x,y \in [0,1]
\end{equation}
with Jacobian $y$
we can rewrite \eqref{eqn:I123} in a form where all integrations range from $0$ to $1$,
\begin{align}\label{eqn:IAIB}
I_A&= c_2s t  \int_0^1 d^3s dx dy  \frac{(x \bar{x} \bar{y})^{\frac{d-2}{2}}}{\Delta_y^{d-1}} \,, \\ \nonumber
I_B&= - 2 (d-1) c_2 s t (s+t) \int_0^1 d^3s dx dy  \frac{(x \bar{y})^{\frac{d}{2}} \bar{x}^{\frac{d-2}{2}} \bar{s}_1 s_3}{\Delta_y^{d}} \,,
\end{align}
where
\begin{align}
\Delta_y &= - \left( s x \bar{s}_1 (\bar{s}_3 \bar{y} + s_2 \bar{x}y) + t \bar{y}s_3(xs_1+\bar{s}_2 \bar{x} ) \right) \,,
\end{align}
and $I_{321}=I_A + I_B$. 
The integral $I_A$ is divergent as $\bar{s}_1, s_3\rightarrow 0$, see also fig. \ref{fig:divergence-vertex}

\subsubsection{Numerical evaluation using the Mellin-Barnes method}
%We choose the Mellin Barnes technique, since it is ideally suited for an automatical extraction of 
%divergences and numerical evaluation of their coefficients, which in particular pays out for the 
%evaluation of the special conformal Ward identies. 
In this section, we switch from the Feynman parametrization in equation \eqref{eqn:IAIB } 
to a Mellin-Barnes representation, as the latter is very convenient to perform a systematic
expansion in $\epsilon$.
An introduction to the Mellin Barnes technique 
can be found in \cite{smirnov2006feynman}. 
%Here we illustrate the technique for the much simpler 
%case of the vertex diagram.\\

In the first step the sum in the denominator is transformed into an integral over a product of terms.  
Since the denominator in \eqref{eqn:I123} consists of a sum of four terms, we will introduce 3 Mellin parameters $z_1,z_2,z_3$.
By repeated use of the Mellin-Barnes representation
\begin{equation}
\frac{1}{(X+Y)^\lambda}= \frac{1}{\Gamma(\lambda)}\frac{1}{2 \pi i} \int_{\beta - i \infty}^{\beta + i \infty}\frac{Y^z}{X^{\lambda +z}}  \Gamma(z+\lambda) \Gamma(-z)dz\,,
\end{equation}
where $- \text{Re} (\lambda) < \beta < 0$, one obtains
\begin{equation}\label{3-fold-MB}
\frac{(2 \pi i)^3 \Gamma(\lambda)}{(a+b+c+d)^\lambda}=\int dz_{1,2,3} a^{z_1}b^{z_2}c^{z_3}d^{-\lambda-z_1-z_2-z_3}\Gamma(-z_1)\Gamma(-z_2)\Gamma(-z_3)\Gamma(\lambda +z_1+z_2+z_3)
\end{equation}
where the real parts $\beta_i$ of the integration contour have to be chosen such that  the arguments in all $\Gamma$ functions have positive real part.
Applying \eqref{3-fold-MB} to the denominator of $I_A$ \eqref{eqn:IAIB}, we can rewrite $I_A$ as
\begin{align}
 I_A&= \frac{c_2}{\Gamma(d-1)}\int d\tilde{z}_{1,2,3} \Gamma(-z_1)\Gamma(-z_2)\Gamma(-z_3)\Gamma(d-1 +z_1+z_2+z_3) (-s)^{z_1+z_2+1}(-t)^{-d - z_1-z_2+2} \\ \nonumber
&\phantom{=} \int d^3s dx dy  s_1^{z_3}\bar{s}_1^{z_1+z_2} s_2^{z_2} \bar{s}_2^{-z_1-z_2-z_3-d+1}s_3^{-z_1-z_2-d+1}\bar{s}_3^{z_1} x^{z_1+z_2+z_3+d/2-1} \bar{x}^{-z_1-z_3-d/2} y^{z_2}\bar{y}^{-z_2-d/2} 
\end{align}
where $d\tilde{z} = (2 \pi i)^{-1}dz$. 
The integrals over $s_1,s_2,s_3,x,y$  can be carried out using 
\begin{equation}
\int_0^1 s_i^{a-1}(1-s_i)^{b-1}dt %= B(a,b) 
={\Gamma(a)\Gamma(b)}/{\Gamma(a+b)}\,,
\end{equation}
and we arrive at 
% an expression that contains only $\Gamma$ functions and is very convenient for the numerical extraction of divergent terms:
\begin{align}\label{eqn:I_Aind3-2eps}
 I_A&= \frac{c_2}{\Gamma(d-1)} \int d\tilde{z}_{1,2,3}{(-s)}^{z_1+z_2+1}{(-t)}^{-d - z_1-z_2+2} \Gamma(-z_1)\Gamma(-z_2)\Gamma(-z_3) \\ \nonumber 
&\times   \Gamma (z_1+1) \Gamma (z_2+1) \Gamma (z_3+1)  \Gamma(d-1 +z_1+z_2+z_3)\Gamma \left(-{d}/{2}-z_2+1\right) \Gamma (z_1+z_2+1) \\ \nonumber
&\times \Gamma(-d-z_1-z_2+2) \Gamma \left(-{d}/{2}-z_1-z_3+1\right) \Gamma (-d-z_1-z_2-z_3+2) \Gamma
  \left({d}/{2}+z_1+z_2+z_3\right)\\ \nonumber
&\times \left[\Gamma \left(2-{d}/{2}\right) \Gamma (-d-z_2+3) \Gamma (-d-z_1-z_3+3)   \Gamma (z_1+z_2+z_3+2)\right]^{-1}\,.
\end{align}
Recall that we investigate the kinematical region where $s,t<0$.

One can see that this integral is divergent as $\epsilon \rightarrow 0$ by noticing that 
for $\epsilon =0$ it is impossible
to choose the integration contours such that all poles of $\Gamma(...+z_1)$ are to the left of the integration contour 
and all poles of $\Gamma(...-z_1)$ are to the right of the contour. \footnote{This is necessary in order for the previous steps to be well-defined.} 
The reason is that the poles of $\Gamma(z_1 + z_2 +1)$ and $\Gamma (-d-z_1-z_2-z_3+2) =\Gamma(-z_1 - z_2 -1+2 \epsilon)$ ``glue together'' at $z_1=-z_2-1$ for $\epsilon=0$. 
However, one can find allowed contours for $\epsilon \neq 0$.

By shifting the contour left to the pole at $z_1=-z_2-1$ we pick up a residue. 
The factor of $\Gamma(-z_1 - z_2 -1+2 \epsilon)$ evaluated at the residue, results in a divergent 
factor of $\Gamma(2 \epsilon)$. The remaining integral over the shifted contour yields a finite contribution. 

The steps of shifting contours and taking residues have been automatised in \cite{Czakon:2005rk} 
and we used this package to systematically extract the pole terms.
%Feeding the package with
Applying this procedure to \eqref{eqn:I_Aind3-2eps} and expanding in $\epsilon$ yields 3 integrals:
\begin{align}\label{eqn:intsA1}
I_A^{(1)} &= c_2 \int \frac{dz_1}{2 \pi i}\left(\frac{1}{\epsilon}+2 \log(-s) - g_1(z_1)\right) f_1(z_1) + \mathcal{O}(\epsilon)\\ \label{eqn:intsA2}
 I_A^{(2)} &= c_2   \int \frac{dz_1 dz_3}{(2 \pi i)^2}\left(\frac{1}{\epsilon}+2 \log(-s) - g_2(z_1,z_3)\right) f_2(z_1,z_3) + \mathcal{O}(\epsilon)\\ \label{eqn:intsA3}
 I_A^{(3)} &= c_2  \int \frac{dz_1dz_2 dz_3}{(2 \pi i)^3} \left(\frac{s}{t}\right)^{1+z_1+z_2}f_3(z_1,z_2,z_3)
\end{align}
We do not specify the values of the real parts $\beta_i$ as well as the functions $f_i,g_i$ here, which are lengthy expressions of products of $\Gamma$ functions and can be obtained automatically by expanding   \eqref{eqn:I_Aind3-2eps} with the help of \cite{Czakon:2005rk}.
Adding up the divergent part of \eqref{eqn:intsA1} and \eqref{eqn:intsA2} we get
\begin{equation}
I^{\text{vertex}}_{\text{div}}= \frac{a_1}{\epsilon} c_2 
\end{equation}
where by numerical integration one finds
\begin{align}
a_1 = \int \frac{dz_1}{2 \pi i}f_1(z_1) +\int \frac{dz_1dz_3}{(2 \pi i)^2}f_2(z_1,z_3)
= 8.710344 \pm 10^{-6}
\approx 4 \pi \ln(2)=8.710344361\ldots
\end{align}
accurately approximated by our analytic guess.
Further numerical evaluation of the finite part of $I_A^{(1)}$, $I_A^{(2)}$, $I_A^{(3)}$ yields
\begin{align}
I_{A,\text{finite}} = c_2 \left( a_1 \ln (-s) + a_1 \ln (-t) +a_2 \ln^2 \left(\frac{s}{t}\right) + a_4 \right)\,.
\end{align}
where $a_1$ has the same value as above and 
\begin{align}
a_2 = - 0.84 \pm 0.01\, \qquad  a_4 = -14.375216465 \pm 10^{-9}\,.
\end{align}
Numerical analysis for $I_B$ suggests
\begin{align}
I_B=c_2 \left(a_3\ln^2\left(\frac{s}{t}\right) + a_5 \right)
\end{align}
where
\begin{align}
a_3=3.97\pm0.01\,,\qquad a_5 = 40.620843911 \pm 10^{-9}\,.
\end{align}
Adding up $a_2$, $a_3$ we obtain
\begin{equation}
a_2+a_3=3.136\pm 0.02 \approx \pi\,.
\end{equation}
Thus suggesting
\begin{align}
I_A+I_B&=c_2  \left(\frac{a_1}{\epsilon}+a_1\left(\ln(-s)+\ln(-t) \right) + (a_2+a_3) \ln \left(\frac{s}{t}\right) + (a_4+a_5) + \mathcal{O}(\epsilon) \right)\\ \nonumber
&\approx c_2 \pi \left(\frac{4 \ln(2)}{\epsilon}+4 \ln(2) \left(\ln(-s)+\ln(-t) \right) + \ln^2 \left(\frac{s}{t}\right) + a_6 + \mathcal{O}(\epsilon) \right)
\end{align}
where
\begin{equation}\label{eqn:a6}
a_6=(a_4+a_5)/\pi= 8.354242685 \pm 2\cdot 10^{-9}\,.
\end{equation}
The constant fits the value $a_6 \approx -\frac{2}{3}\pi^2 + 16 \ln(2) + 8 \ln^2 (2) =8.35424273...$ . The result can be rewritten in the form
\begin{align}
I_A+I_B
&\approx c_2 \pi \left(2 \ln(2)\frac{ ((-s) ^{2\epsilon}+(-t)^{2\epsilon})}{\epsilon}  + \ln^2 \left(\frac{s}{t}\right) + a_6 + \mathcal{O}(\epsilon) \right)
\end{align}

\subsection{Gauge field and ghost loops}\label{app:gluon-and-ghost-loops}
It is well known \cite{Chen:1992ee}, that the contributions of ghost and gauge field loops to the gauge field self energy cancel. We briefly review the cancellation of the gauge field and ghost loop corrections in the Wilson loop, since from this it is easy to see, how the cancellation for the insertions in the conformal Ward identities takes place.  
%The relevant part of the action is
%\begin{align}
% S=  \frac{k}{4\pi}\int d^dx\tr\left(A_\mu \partial_\nu A_\rho - \frac{2}{3} i A_\mu A_\nu A_%rho\right)- \frac{k}{4\pi}\int d^dx\tr\left( \partial^\mu \bar{c}\, D_\mu c \right)+...
%\end{align}
%where
%\begin{equation}
%D_\mu c=\partial_\mu c + i [A^\mu,c] \,.
%\end{equation}
\subsubsection{Gauge field loop}
The gauge field loop-diagram arises at second order in perturbation theory
\begin{align}
\langle W_n \rangle^\text{gluon-loop}&= \frac{1}{N}\langle \tr(-\oint dz_i^\mu dz_j^\nu A_\mu A_\nu ) \left(-\frac{1}{2!}\right)\left( \frac{k}{4\pi}  \int d^dx\epsilon^{\alpha\beta\gamma}  \tr(\frac{2}{3}iA_\alpha A_\beta A_\gamma)\right)^2 \rangle \\ \nonumber
&=c_8 \left(\frac{2}{3}\right)^2\oint dz_i^\mu dz_j^\nu \int d^dx d^dy  \epsilon^{\alpha\beta\gamma} \epsilon^{\delta\sigma\tau} \langle \tr(A_\mu A_\nu) \tr (A_\alpha A_\beta A_\gamma)(x)\tr (A_\delta A_\sigma A_\tau)(y) \rangle 
\end{align}
where 
\begin{equation}
c_8 = -\frac{1}{N}\frac{1}{2} \left(\frac{k}{4\pi} \right)^2\,.
\end{equation}
Taking into account that $A_\mu$, $A_\nu$ give 3 identical contractions with one of the vertex terms and that we can contract them either with the $x$- or $y$-dependent vertex, we get a symmetry factor of $3\cdot 3\cdot 2$. The remaining contractions of the gauge fields are dictated by taking into account only planar diagrams. Thus we get
\begin{align}\label{eqn:gluon-loop}
\langle W_n \rangle^\text{gluon-loop}&= 8 c_8  \oint dz_i^\mu dz_j^\nu \int d^dx d^dy \epsilon^{\alpha\beta\gamma} \epsilon^{\delta\sigma\tau} \langle A_\mu A_\alpha \rangle \langle A_\nu A_\delta \rangle \langle A_\beta A_\tau \rangle \langle A_\gamma A_\sigma \rangle \,.
\end{align}
To proceed, we recall the relation \eqref{eqn:ghost-gluon-propagator-relation} between 
gauge field and ghost propagator and write 
\begin{align}
\epsilon^{\alpha\beta\gamma} \epsilon^{\delta\sigma\tau}  \langle A_\beta A_\tau \rangle \langle A_\gamma A_\sigma \rangle 
&= \frac{1}{4}\, {\epsilon^{\alpha\beta\gamma} \epsilon^{\delta\sigma\tau} \epsilon_{\beta\tau\kappa} \epsilon_{\gamma\sigma\rho}}\partial^\rho_{x} \langle c(x) \bar{c}(y) \rangle \partial^\kappa_{x} \langle c(x) \bar{c}(y) \rangle \nn\\
&= \frac{1}{2} \partial^\rho_{x} \langle c(x) \bar{c}(y) \rangle \partial^\kappa_{y} \langle c(x) \bar{c}(y) \rangle
\end{align}
where we used $
\epsilon^{\alpha\beta\gamma} \epsilon^{\delta\sigma\tau} \epsilon_{\beta\tau\kappa}\epsilon_{\gamma\sigma\rho} ={- \left(\eta^\alpha_\kappa \eta^\delta_\rho+\eta^\alpha_\rho \eta^\delta_\kappa \right)}$ and
$\partial_{x} F(x-y)= - \partial_{y} F(x-y)$ in the last step.

\subsubsection{Ghost loop}
The ghost loop diagram arises from contraction of the second order perturbation theory expansion of the gauge-field-ghost vertex term
\begin{align}
\langle W_n \rangle^\text{ghost loop}&= 
\frac{1}{N}\langle \tr(-\oint dz_i^\mu dz_j^\nu A_\mu A_\nu ) \left(-\frac{1}{2!}\right)\left( \frac{k}{4\pi}  \int d^dx  \tr(\partial^\mu \bar{c}\,i[A_\mu,c] ) \right)^2 \rangle \\ \nonumber
&= 2c_8 \oint dz_i^\mu dz_j^\nu \int d^dx d^dy  \langle \tr(A_\mu A_\nu) \tr(\partial^\rho_{x} \bar{c} A_\rho c) \tr(\partial^\sigma_{y} \bar{c} A_\sigma c)  \rangle 
\end{align}
where $c_8$ ist the same as defined above and the factor of $2$ is due to 
the fact that the evaluation of the first line yields two identical planar diagrams 
that are kept and two identical non-planar diagrams that we drop. 
Contracting $A_\mu$ either with the $x$- or $y$-dependent vertex, we get a symmetry factor of 2. 
There is only one way for the remaining contractions and thus we get
\begin{align}\label{eqn:ghost-loop}
\langle W_n \rangle^\text{ghost loop}&=-4 c_8\,  \oint dz_i^\mu dz_j^\nu \int d^dx d^dy \langle A_\mu A_\sigma \rangle  \langle A_\nu A_\rho \rangle \partial_x^\rho \langle c(y) \bar{c}(x) \rangle \partial_y^\sigma \langle c(x) \bar{c}(y) \rangle
\end{align}
where a factor of $-1$ due to the anti-commuting ghost fields in the loop was taken into account.
Summing up \eqref{eqn:gluon-loop} and \eqref{eqn:ghost-loop} we get
\begin{align} \langle W_n \rangle^\text{gauge field loop}+\langle W_n \rangle^\text{ghost loop}=0\,.
\end{align}
The same relation \eqref{eqn:ghost-gluon-propagator-relation} can be used to show the vanishing for the dilatation and special conformal Ward identities.

\subsection{Conformal Ward identity}\label{app:conf-wi-two-loop}

\subsubsection{Insertion of the interaction term} \label{app:conf-wi-insertion-interaction-term}
We can rewrite \eqref{eqn:I321prime} as
\begin{align}
I_{321}^\prime&=\frac{1}{(d-2)^2}\int d^3s_{1,2,3} \epsilon(p_1,p_2,\partial_{z_1})\epsilon(p_3,p_2,\partial_{z_2}) \int d^dx  \frac{(x+z_2)^\nu}{|x|^{d}|x-z_{12}|^{d-2}|x-z_{32}|^{d-2}}\,.
\end{align}
Introducing Feynman parameters, changing the integration variable to $l=x-\beta_1 z_{12}-\beta_3 z_{32}$, using the same notation as in app. \ref{app:Vertex}, integrating over $l$ 
%\begin{align}
%I_{321}^\prime&=\frac{(-1)^{\frac{3d}{2}-2}}{(d-2)^2}\int d^3s_{1,2,3} %\epsilon(p_1,p_2,\partial_{z_1})\epsilon(p_3,p_2,\partial_{z_2}) \int d[\beta]_3\int %d^dl  \frac{(l + \beta_1 z_1+ \beta_2 z_2+ \beta_3 %z_3)^\nu}{(l^2-\Delta)^{\frac{3d}{2}-2}}\\ \nonumber
%&= \frac{1}{(d-2)^2}\int d^3s_{1,2,3} %\epsilon(p_1,p_2,\partial_{z_1})\epsilon(p_3,p_2,\partial_{z_2}) \int d[\beta]_3 %(\beta_1 z_1+ \beta_2 z_2+ \beta_3 z_3)^\nu \frac{c_1}{\Delta^{d-2}}
%\end{align}
and evaluating the action of the derivatives yields
\begin{align}
I_{321}^\prime &=\frac{c_1}{(d-2)^2}\int d^3s_{1,2,3} \int d[\beta]_3\Big( (\beta_1 z_1+ \beta_2 z_2+ \beta_3 z_3)^\nu  \epsilon(p_1,p_2,\partial_{z_1})\epsilon(p_3,p_2,\partial_{z_2}) \frac{1}{\Delta^{d-2}}
\\ \nonumber 
&\phantom{=}\quad \qquad \qquad \qquad\qquad\qquad +\epsilon(p_1,p_2,p_3)2 \beta_1 \beta_3 \epsilon_{\alpha\beta}^{~~~\nu}p_2^\beta \left(\beta_1p_1^\alpha \bar{s}_1+\beta_3 p_3^\alpha s_3 \right)\frac{(2-d)}{\Delta^{d-1}} \Big)\,,
\end{align}
The last term can be shown to be finite and the first term is very similar to the vertex diagram. Evaluation of the derivatives as in \ref{app:Vertex} yields
\begin{align}\label{eqn:I123}
I_{321}^\prime &= c_2  s t  \int_0^1 d^3s_{1,2,3} d^3\beta_{1,2,3}(\beta_1\beta_2\beta_3)^{(d-2)/2} \delta(\sum_i \beta_i-1)\\ \nonumber
&\phantom{=}\qquad \qquad (\beta_1 z_1+ \beta_2 z_2+ \beta_3 z_3)^\nu \left( \frac{ 1}{\Delta^{d-1}}- 2\frac{(d-1)}{\Delta^{d}}\beta_1\beta_3 \bar{s}_1 s_3 (s+t)  \right) + \text{finite}\,.
\end{align}
It can be shown, e.g. using the Mellin Barnes technique as in \ref{app:Vertex}, that all divergent contributions are due to the first term. We have the following divergent contributions:
\begin{align}
st \int d^3s d[\beta]_3 \beta_1\beta_2\beta_3 \left(\frac{\beta_1 z_1}{\Delta^{d-1}}\right)&= \frac{1}{\epsilon} \, {a\, x_2^\nu } + \mathcal{O}(\epsilon^0)\,,\\ \nonumber
st \int d^3s d[\beta]_3 \beta_1\beta_2\beta_3 \left(\frac{\beta_2 z_2}{\Delta^{d-1}}\right)&= \frac{1}{\epsilon}\, {b \, (x_2^\nu+x_3^\nu) } + \mathcal{O}(\epsilon^0)\,, \\ \nonumber
st\int d^3s d[\beta]_3 \beta_1\beta_2\beta_3 \left(\frac{\beta_3 z_3}{\Delta^{d-1}}\right)&= \frac{1}{\epsilon}\left(a\, x_3^\nu  \right) + \mathcal{O}(\epsilon^0)\,.
\end{align}
Numerical evaluation yields
\begin{align}
a= 1.8562\pm 0.0001 \,,\quad
b= 2.4989 \pm 0.0001\,.
\end{align}
To good accuracy we find
\begin{equation}
a+b=4.35517 \pm 0.0002 \approx 2 \pi \ln(2) =4.35517...\,.
\end{equation}
Summarising $I_{321}^\prime$ then reads
\begin{equation}
I_{321}^\prime =  \frac{c_2}{\epsilon} (a+b)(x_2+x_3)^\nu + \mathcal{O}(\epsilon^0)
\approx  \frac{2 \pi i\ln(2)}{\epsilon}(x_2+x_3)^\nu+ \mathcal{O}(\epsilon^0)\,.
\end{equation}

\subsubsection{Insertion of the kinetic term into the vertex diagram}\label{app:conf-wi-insertion-kinetic-term-in-vertex}
For the kinetic insertion into the vertex diagram we have \eqref{eqn:kinetic-insertion-in-vertex}
\begin{align}\langle \mathcal{L}(x) W_4 \rangle^{(2)}_{\text{(c)}} 
&=  \underbrace{\left( \frac{N}{k}\right)^2 \frac{i}{8 \pi^2} \left( \frac{\Gamma\left(\frac{d}{2}\right)}{\pi^{\frac{d-2}{2}}}\right)^4}_{=: c_3} \int d^dw  \oint dz_{i,j,k}^{\mu\nu\rho} \epsilon^{\delta\sigma\tau} I_{\nu\sigma}\, G_{\mu\tau}(z_i-w) G_{\rho\delta}(z_k-w) \\ \nonumber
& + \text{cyclic}(\mu,\nu,\rho;z_i,z_j,z_k)
\end{align}
where $G_{\mu\nu}(x-y)=\epsilon_{\mu\nu\rho} \frac{(x-y)^\rho}{\left(-(x-y)^2\right)^{\frac{d}{2}}}$ and
\begin{align}\label{eqn:integration-yields-prop}
I_{\nu\sigma}(x- z_j,x-w) &=\epsilon^{\alpha\beta\gamma}\left(G_{\alpha\nu}(x-z_j) \partial_\beta^{(x)} G_{\gamma\sigma}(x-w) + G_{\alpha\sigma}(x-w) \partial_\beta^{(x)} G_{\gamma\nu}(x-z_j) \rangle \right) \,.
\end{align} 
and the two other contractions are contained in cyclic$(\mu,\nu,\rho;z_i,z_j,z_k)$.
For the dilatation Ward identity the integration over $x$ can be performed by introducing two Feynman parameters. The result simply yields a propagator
\begin{align}
\int d^dx\, I_{\nu\sigma}
&= - \frac{4  \pi^{\frac{d}{2}}}{\Gamma\left(\frac{d}{2}\right)} \epsilon_{\nu\sigma\varphi}  \frac{(z_j-w)^\varphi}{(-(z_j-w)^2)^{\frac{d}{2}}} = - \frac{4   \pi^{\frac{d}{2}}}{\Gamma\left(\frac{d}{2}\right)} G_{\nu\sigma}(z_j-w)\,,
\end{align}
In the case of the special conformal Ward identity the integration is a little more involved. The integral over $d^dx$ can be solved by introducing Feynman parameters and after some algebra one finds
\begin{align}
\int d^d x \,x^\lambda  \langle \mathcal{L}(x) W_4\rangle^{(2)}_{(c)} &= 2 c_3 c_4 \oint dz_{i,j,k}^{\mu,\nu,\rho}\int d^dw  \epsilon^{\delta\sigma\tau}\epsilon^{\alpha\beta\gamma}\epsilon_{\alpha\nu\xi} \epsilon_{\gamma\sigma\varphi} \\ \nonumber
&\phantom{=} \partial^\xi \left((2 \eta^{\varphi\lambda}\partial_\beta + \eta_\beta^{\lambda}\partial^\varphi) \left(\frac{1}{((z_j-w)^2)^{\frac{d}{2}-2}}\right)- \partial^\varphi \partial_\beta\left(\frac{(z_j+w)^{\lambda}}{((z_j-w)^2)^{\frac{d}{2}-2}}\right)\right) \\ \nonumber
&\phantom{=} G_{\rho\delta} G_{\mu\tau} + \text{cyclic} (\mu,\nu,\rho; z_i,z_j,z_k)\,,
\end{align}
where all derivatives are taken with respect to $z_j$ and $c_4$ is a constant obtained through integration over $d^dx$. Inserting the propagators we can write this in a form convenient to solve the integral over $d^dw$
\begin{align}
&2 c_3 c_4 \oint dz_{i,j,k}^{\mu,\nu,\rho} \epsilon^{\delta\sigma\tau}\epsilon^{\alpha\beta\gamma}\epsilon_{\alpha\nu\xi} \epsilon_{\gamma\sigma\varphi}\epsilon_{\rho \delta \chi} \epsilon_{\mu \tau \theta} \partial^\theta_i \partial^\xi_j \partial^\chi_k \left((2 \eta^{\varphi\lambda}\partial_{j,\beta} + \eta_\beta^{\lambda}\partial^\varphi_j) J_j- \partial^\varphi_j \partial_{\beta,j} J_j^\lambda \right) 
+ \text{cyclic}
\end{align}
where the integrals read
\begin{align}
J_j &= \frac{1}{ (2-d)^2} \int d^dw \frac{1}{((z_j-w)^2)^{\frac{d}{2}-2} ((z_i-w)^2)^{\frac{d}{2}-1} ((z_k-w)^2)^{\frac{d}{2}-1}}  \\ \nonumber
J_j^{\lambda}&= \frac{1}{ (2-d)^2} \int d^dw \frac{(z_j+w)^{\lambda}}{((z_j-w)^2)^{\frac{d}{2}-2} ((z_i-w)^2)^{\frac{d}{2}-1} ((z_k-w)^2)^{\frac{d}{2}-1}}
\end{align}
The integrations over $w$ can be performed by introducing three Feynman parameters $\beta_i$ and we get
\begin{align}
J_j &=  c_5\int d[\beta]_{3,j} \left(\frac{1}{\Delta}\right)^{d-4}\,, \qquad
J_j^\lambda = c_5\int d[\beta]_{3,j}  (z_j + \sum_i \beta_i z_i )^\lambda \left(\frac{1}{\Delta}\right)^{d-4}
\end{align}
where
\begin{align}\label{eqn:feynman-measure}
\int d[\beta]_{3,j} &= \int_0^1 d\beta_i d\beta_j d\beta_k \delta\left(\sum_i \beta_i  -1\right) \left(\beta_i \beta_j \beta_k \right)^{\frac{d}{2}-2} \beta_j^{-1}
\end{align}
and $c_5$ is a constant obtained by integrating over $w$, the product $c_3 c_4 c_5 $ is explicitly given below.
The expression for $\Delta$ is the same as in \eqref{eqn:Delta-denominator-vertex}
For the cyclic permutations we can use the same expression, replacing the measure with $d[\beta]_{3,i}$ respectively $d[\beta]_{3,k}$ , i.e. exchanging $\beta_j^{-1}$ with $\beta_i^{-1}$ respectively $ \beta_k^{-1}$ in \eqref{eqn:feynman-measure}. \\

All three contributions can then be written as
\begin{align}
2 c_3 c_4 c_5 \oint dz_{i,j,k}^{\mu,\nu,\rho} \epsilon^{\delta\sigma\tau}\epsilon^{\alpha\beta\gamma}\epsilon_{\gamma\sigma\varphi}
& \Big( \epsilon_{\alpha\mu\xi} \epsilon_{\nu \delta \chi} \epsilon_{\rho \tau \theta} \partial^\theta_k \partial^\xi_i \partial^\chi_j \left((2 \eta^{\varphi\lambda}\partial_{i,\beta} + \eta_\beta^{\lambda}\partial^\varphi_i) J_i- \partial^\varphi_i \partial_{\beta,i} J_i^\lambda \right) \\ \nonumber
& \phantom{\Big(}\epsilon_{\alpha\nu\xi} \epsilon_{\rho \delta \chi} \epsilon_{\mu \tau \theta} \partial^\theta_i \partial^\xi_j \partial^\chi_k \left((2 \eta^{\varphi\lambda}\partial_{j,\beta} + \eta_\beta^{\lambda}\partial^\varphi_j) J_j- \partial^\varphi_j \partial_{\beta,j} J_j^\lambda \right) \\ \nonumber
& \phantom{\Big(} \epsilon_{\alpha\rho\xi} \epsilon_{\mu \delta \chi} \epsilon_{\nu \tau \theta} \partial^\theta_j \partial^\xi_k \partial^\chi_i \left((2 \eta^{\varphi\lambda}\partial_{k,\beta} + \eta_\beta^{\lambda}\partial^\varphi_k) J_k- \partial^\varphi_k \partial_{\beta,k} J_k^\lambda \right) 
\Big)
\end{align}
and where 
\begin{align}
2 c_3 c_4 c_5 = i\frac{ \pi^{2-d}}{128} \left(\frac{N}{k}\right)^2 \Gamma(d-4)
\end{align}
We can evaluate the derivatives and contractions with the computer and find that the non-vanishing contributions have the structure
\begin{align}
&\phantom{=} \int d^dx\,x^\lambda \langle \mathcal{L}(x) W_4 \rangle^{\text{vertex-insertion}}=\left(\frac{N}{k}\right)^2 \sum_{i>j>k}\int_0^1 ds_{i,j,k}\int d[\beta]_3\left(I_{ijk,-d-1}^\lambda+I_{ijk,-d}^\lambda+I_{ijk,-d+1}^\lambda \right)
\end{align}
where $I_{ijk,p}$ are lengthy terms proportional to $1/\Delta^p$.

For the conformal Ward identity we are only interested in the divergent part of the above quantities, which can be automatically extracted with the Mellin-Barnes technique. We find that all terms vanish except for $i\neq j \neq k$. Specialising to the case $i=3, j=2, k=1$ we find
\begin{align}
\int I_{312,-d-1}^\lambda&= \mathcal{O}(\epsilon^0)\,,\\ \nonumber
\int I_{312,-d}^\lambda&= \frac{i}{\epsilon} ( a_1 x_2^\lambda+a_2 x_3^\lambda)+ \mathcal{O}(\epsilon^0)\,, \\ \nonumber
\int I_{321,-d+1}^\lambda&= \frac{i}{\epsilon} ( b_1 x_2^\lambda+b_2 x_3^\lambda)+  \mathcal{O}(\epsilon^0)\,.
\end{align}
Numerical evaluation of the integrals yields
\begin{align}
a_1 &= 0.3465735 \pm 10^{-6}  \approx \frac{1}{2}\ln(2) = 0.3465735...\,, \\ \nonumber
a_2 &= 0.3465735 \pm 8\cdot 10^{-7}  \approx \frac{1}{2}\ln(2) = 0.3465735...\,, \\ \nonumber
b_1 &=-0.8664339 \pm 14\cdot 10^{-7}\approx -\frac{5}{4}\ln(2) = -0.86643397...\,, \\ \nonumber
b_2 &=-0.8664339 \pm 10^{-6} \approx -\frac{5}{4}\ln(2) = -0.86643397...\,. 
\end{align}
Adding up the results, summing over all four diagrams and taking into account the corresponding prefactors we get
\begin{align}
&\phantom{=} \int d^dx\,x^\lambda \langle \mathcal{L}(x) W_4 \rangle^{(2)}_{(c)} \approx -i\frac{3}{4} \frac{\ln(2)}{\epsilon}\left(\sum_i x_i^\lambda\right)+  \mathcal{O}(\epsilon^0)\,.
\end{align}

\section{One loop gauge field propagator in ABJM theory}\label{sec:one-loop-correction-propagator}
Here we review the calculation of the one-loop correction to the gauge field propagator, see also \cite{Drukker:2008zx}. We have fermionic and bosonic contributions in the loop and thus
\begin{equation}
G_{\mu\nu}^{(1)}(p) =G_{\mu\nu}^{(F,1)}(p) +G_{\mu\nu}^{(B,1)}(p) 
\end{equation}
where 
\begin{equation}\label{eqn:Gmunu}
G_{\mu\nu}^{(1)}(p) = \left(\frac{2\pi}{k}\right)^2 \frac{\epsilon_{\mu\rho\kappa}p^\kappa}{p^2}\left( \Pi_{\rho\lambda}^{(B)}(p) + \Pi_{\rho\lambda}^{(F)}(p) \right) \frac{\epsilon_{\lambda\nu\delta}p^\delta}{p^2}
\end{equation}
and
\begin{align}
 \Pi_{\mu\nu}^{(B)}(p) &=+N \delta^I_I \mu^{2\epsilon}\int \frac{ d^dk}{(2\pi)^d} \frac{(2k+p)_\mu(2k+p)_\nu}{k^2(p+k)^2} \\ \nonumber
 \Pi_{\mu\nu}^{(F)}(p) &= -N \delta^I_I \mu^{2\epsilon}\int \frac{ d^dk}{(2\pi)^d} \frac{\tr\left(\gamma_\mu(\pslash +\kslash) \gamma_\nu \kslash \right)}{k^2(p+k)^2}\,.
\end{align}
We use the DRED scheme for Dirac matrix operations as well as for Levi-Civita tensor contractions, i.e. we work in strictly $d=3$ to obtain scalar integrands and only then continue the loop momenta to d-dimensional space to perform the integrals in d dimensions. This scheme has been shown to respect the Slavnov-Taylor identities up to two loop order in \cite{Chen:1992ee}.

 Then we have
\begin{equation}
\tr\left(\gamma_\mu(\pslash +\kslash) \gamma_\nu \kslash \right) = 2\left(-\eta_{\mu\nu} (p+k) \cdot k+ 2k_\mu k_\nu+ p_\mu k_\nu + p_\mu k_\nu \right)\,.
\end{equation}
The last two terms can be dropped, since they vanish when contracted with \eqref{eqn:Gmunu}. The same is true for terms proportional to $p_\mu, p_\nu$ in the bosonic term.

Summing up all remaining terms we get
\begin{equation}
+N \delta^I_I \mu^{2\epsilon} 2 \eta_{\mu\nu}\int \frac{ d^dk}{(2\pi)^d} \frac{ k \cdot (p+k)}{k^2(p+k)^2}
\end{equation}
Introducing Feynman parameters, we have
\begin{align}
+N \delta^I_I \mu^{2\epsilon}2 \eta_{\mu\nu} \int_0^1 d\alpha \int \frac{ d^dk}{(2\pi)^d}  \frac{ k \cdot (p+k)}{[(k+\bar{\alpha}p)^2 - \Delta]^2}
\end{align}
where $\Delta= - \alpha \bar{\alpha}p^2$. Then, we shift $k=l-\bar{\alpha} p$ and drop terms linear in $l_\mu$
\begin{align}
+N \delta^I_I \mu^{2\epsilon}2 \eta_{\mu\nu} \int_0^1 d\alpha \int \frac{ d^dl}{(2\pi)^d}  \frac{ l^2-\alpha \bar{\alpha} p^2}{[l^2 - \Delta]^2}\,.
\end{align}
Using the standard integrals
\begin{equation}
\int \frac{ d^dl}{(2\pi)^d} \frac{l^2}{[l^2-\Delta]^2}= -\frac{i}{(4 \pi)^{\frac{d}{2}}} \frac{d}{2} \frac{\Gamma(1-\frac{d}{2})}{\left( \Delta \right)^{1-\frac{d}{2}}}\, \qquad \int \frac{ d^dl}{(2\pi)^d} \frac{1}{[l^2-\Delta]^2}= \frac{i}{(4 \pi)^{\frac{d}{2}}}  \frac{\Gamma(2-\frac{d}{2})}{\left( \Delta \right)^{2-\frac{d}{2}}}\,
\end{equation}
and $\epsilon_{\lambda\kappa\mu}\epsilon_{\lambda\nu\delta} = \eta_{\kappa\nu}\eta_{\mu\delta}-\eta_{\kappa\delta}\eta_{\mu\nu}$ we get
\begin{align}\label{eqn:Gmunuinserted}
G_{\mu\nu}^{(1)}(p) 
&= \left( N \delta^I_I \mu^{2\epsilon}2  \frac{(-i)}{(4\pi)^{\frac{d}{2}}}  \frac{\Gamma(1-\frac{d}{2}) \Gamma(\frac{d}{2})^2}{\Gamma(d-1)} \right) \left(\frac{2\pi}{k}\right)^2  \frac{1}{{(-p^2)^{3-\frac{d}{2}}}}\left(p_\mu p_\nu - \eta_{\mu\nu}p^2 \right)\,.
\end{align}
The standard formula\footnote{For $\eta_{\mu\nu}= diag\,(1,-1,-1)$}
\begin{align}
\int \frac{d^dp}{(2\pi)^d} \frac{e^{-ipx}}{(-p^2)^k} 
&= i \frac{\Gamma(\frac{d}{2}-k)}{\Gamma(k)} \frac{1}{4^k \pi ^\frac{d}{2}} \frac{1}{(-x^2)^{\frac{d}{2}-k}} 
\end{align}
leads to the Fourier transform of \eqref{eqn:Gmunuinserted}
\begin{align}\label{eqn:one-loop-correctd-gluon-prop.}
G_{\mu\nu}^{(1)}&(x) = \mu^{2\epsilon} \int \frac{d^dp}{(2\pi)^d}  G_{\mu\nu}^{(1)}(p)  e^{-ipx} \\ \nonumber
&= \left(\frac{2\pi}{k}\right)^2  \frac{N \delta^I_I}{8} \frac{\Gamma(1-\frac{d}{2}) \Gamma(\frac{d}{2})^2}{\Gamma(d-1)}   \frac{(\mu^{2\epsilon})^2}{\pi^d} \left(  \frac{\Gamma(d-2)}{\Gamma(2-\frac{d}{2})} \frac{\eta_{\mu\nu}}{(-x^2)^{d-2}} -\partial_\mu \partial_\nu\left( \frac{\Gamma(d-3)}{\Gamma(3-\frac{d}{2})} \frac{1}{4}  \frac{1}{(-x^2)^{d-3}}  \right) \right)
\end{align}

%%%%%%%%%%%%%%%%%%%%%%%%%%%%%%%%%%%%%%%%%%%%%%%%%%%%%%%%%%%%%%%%%%%%%%%%%%%%%%%%
%%%%%%%%%%%%%%%%%%%%%%%%%%%%%%%%%%%%%%%%%%%%%%%%%%%%%%%%%%%%%%%%%%%%%%%%%%%%%%%%
\bibliographystyle{nb.bst}
\bibliography{cswilson}

\end{document}